\documentclass[final,twoside,journal]{IEEEtran}
\usepackage{graphicx}
\usepackage{subfigure}
\usepackage{graphicx}
\usepackage{cite}
\def\vec#1{\mbox{\boldmath $#1$}}
\def\G{\mathit{\Gamma}}
\def\e{\mathrm{e}}

\begin{document}

\title{Stability of an electromagnetically levitated spherical sample in
a set of coaxial circular loops}

\author{J\={a}nis~Priede~and~Gunter~Gerbeth%
\thanks{Manuscript received December 13, 2004; revised February 24, 2005.
This work was supported by Deutsche Forschungsgemeinschaft in frame
of the Collaborative Research Centre SFB 609 and by the European Commission
under grant No. G1MA-CT-2002-04046.}%
\thanks{J. Priede was with Forschungszentrum Rossendorf, P.O. Box 510119,
01314 Dresden, Germany during this work. Presently he is with the
Institute of Physics, University of Latvia, Miera st. 32, LV--2169
Salaspils, Latvia; \protect \\
\indent G. Gerbeth is with Forschungszentrum Rossendorf, MHD Department,
P.O. Box 510119, 01314 Dresden, Germany}}

\markboth{IEEE Transactions on Magnetics,~Vol.~41, No.~6,~June~2005}{Priede and Gerbeth: Stability of an electromagnetically levitated spherical sample}
\pubid{0018--9464/\$20.00~\copyright~2005 IEEE}

\maketitle
\begin{abstract}
This paper presents a theoretical study of oscillatory and rotational
instabilities of a solid spherical body levitated electromagnetically
in axisymmetric coils made of coaxial circular loops. We apply our
previous theory to analyze the static and dynamic stability of the
sample depending on the AC frequency and the position of the sample
in the coils for several simple configurations. An original analytical
approach is introduced employing a gauge transformation for the vector
potential. First we calculate the spring constants which define the
frequency of small-amplitude oscillations. For static stability the
spring constants must be positive. Dynamic instabilities are characterized
by critical AC frequencies which, when exceeded, may result either
in a spin-up or oscillations with increasing amplitude. It is found
that the critical frequencies increase with the non-uniformity of
the field. We show that for a spherically harmonic field the critical
frequency for the spin-up instability in a field of degree $l$ coincides
with the critical frequency for the oscillatory instability in a field
of degree $l+1$.
\end{abstract}

\section{Introduction}

\PARstart{E}{lectromagnetic} levitation melting (ELM) was invented
in the twenties of the last century \cite{Muck-23} whereas its usage
started only at the beginning of the fifties \cite{Okressetal-52}
when high-frequency power generators became available. The basic principle
of ELM is simple: a conducting sample, usually metallic, weighting
from several tens of grams up to several kilograms is placed in a
coil fed by an AC current with a typical frequency ranging from about
$10\, kHz$ up to several $100\, kHz$. The AC magnetic field induces
eddy currents in the sample which, in turn, give rise to two effects.
On one hand, eddy currents interact with those in the coil giving
rise to a Lorentz force that repels the sample from the coil. On the
other hand, the Ohmic dissipation due to the induced currents provides
heating of the sample. In such a way, the sample can be levitated
and also melted, provided that the coil configuration and the current
in it are properly adjusted. ELM is particularly useful for melting
reactive metals with high melting points, for example such as \textit{Ti},
\textit{Zr}, \textit{V}, \textit{Ta}, \textit{Mo}, which often react
with the crucible material and so get polluted by it \cite{Okressetal-52}.
ELM avoids contamination of the melt and allows one to carry out solidification
from deeply under-cooled states which is of interest for certain material
science applications \cite{Egryetal-01}. Besides, ELM is a well-known
method for measurements of material properties of liquid metals used
on ground as well as in space \cite{Egryetal-95,Lohoeferetal-04}.

% needed in second column of first page if using \pubid
\pubidadjcol

The balance of gravity and electromagnetic forces is necessary but
not sufficient for a successful levitation. In addition, the sample
has to be stable at least to perturbations of sufficiently small amplitude.
First, it means that the reaction force due to the displacement of
the sample from its equilibrium position has to act against that displacement.
Otherwise, the equilibrium will be statically unstable and the solid
sample will fall out of the coil or touch it when it is slightly perturbed.
Similarly, a molten sample may leak out of the coil in the result
of surface folding and deformation \cite{HullRote-89}. The static
stability alone may also be insufficient for a successful levitation
because sometimes the sample exhibits overstability. Namely, there
is a restoring force that makes the sample to execute oscillations
with increasing amplitude so that the sample eventually hits the coil
or leaves it \cite{Okressetal-52}. Similarly, sometimes the sample
is observed to spin-up \cite{Okressetal-52,KeissigEssman-79,AbedianHayers-03}.
A purely electromagnetic theory trying to account for such dynamic
instabilities has been proposed in our previous work and applied to
some simple configurations of magnetic fields \cite{PriedeGerbeth-00o,PriedeGerbeth-00r}.
In this study, we apply our previous theory to analyze both the static
and dynamic stability of a spherical sample in more realistic axisymmetric
magnetic systems made of a set of coaxial circular current loops.
Although the basic configuration is axisymmetric, the perturbation
fields are, in general, three-dimensional that renders the problem
mathematically more complicated. To calculate 3D fields for a spherical
sample we use similar analytic techniques as in Refs. \cite{Lohoefer-93,Lohoefer-03}.
However, our approach differs from the previous ones by an original
use of the gauge transformation for the 3D vector potential in order
to satisfy boundary conditions for the induced current. This allows
us to carry out the analysis only in terms of the vector potential
without considering the scalar electrostatic potential. Therefore
our approach is considerably simpler compared to \cite{Lohoefer-93,Lohoefer-03}.

The paper is organized as follows. The stability of small amplitude
oscillations of arbitrary direction is considered in Sec. 2 beginning
with the governing equations and analytic solutions for the vector
potential of a circular current loop in spherical harmonics. In this
section we also derive solutions for the axisymmetric base state and
the perturbation field due to a small displacement. Section 3 presents
governing equations and analytic solution for the spin-up instability.
Numerical results for both instabilities in several simple inductors
are discussed in Section 4 and summarized in Section 5.

\section{Small-amplitude oscillations}

\subsection{Formulation of the problem and governing equations}

Consider a sphere with radius $R$ and conductivity $\sigma$ moving
at velocity $\vec{v}$ in a magnetic field $\vec{B}$ alternating
with circular frequency $\omega$. The induced electric field follows
from the first Maxwell equation as $\vec{E}=-\vec{\nabla}\Phi-\partial_{t}\vec{A},$
where $\Phi$ is the scalar potential of the electric field and $\vec{A}$
is the vector potential of the magnetic field. The density of the
electric current induced in a moving medium is given by Ohm's law
\[
\vec{j}=\sigma(\vec{E}+\vec{v}\times\vec{B})=\sigma(-\vec{\nabla}\Phi-
\partial_{t}\vec{A}+\vec{v}\times\vec{\nabla}\times\vec{A}).
\]
The electric field induced by a translational solid body motion with
$\vec{v}$ being a spatially invariant vector can be represented as
\[
\vec{v}\times\vec{\nabla}\times\vec{A}=\vec{\nabla}(\vec{v}\cdot\vec{A})-
(\vec{v}\cdot\vec{\nabla})\vec{A}.
\]
Assuming that the frequency of the alternating magnetic field is sufficiently
low to neglect the displacement current, the second Maxwell equation
leads to the following advection-diffusion equation 
\begin{equation}
\partial_{t}\vec{A}+(\vec{v}\cdot\vec{\nabla})\vec{A}=
\frac{1}{\mu_{0}\sigma}\nabla^{2}\vec{A}
\label{eq:vecpot}
\end{equation}
where the gauge invariance of $\vec{A}$ has been employed to define
the scalar potential as 
\begin{equation}
\Phi=\vec{v}\cdot\vec{A}-\frac{1}{\mu_{0}\sigma}\vec{\nabla}\cdot\vec{A}.
\label{eq:gauge}
\end{equation}
Boundary conditions at the surface follow from the continuity of
the magnetic field 
\begin{equation}
\left[\vec{A}_{0}\right]_{S}=\left[\partial_{n}\vec{A}_{0}\right]_{S}=0
\label{eq:bcnd-1}
\end{equation}
where $\left[f\right]_{S}$ denotes a jump of the quantity $f$ across
the boundary $S$; $\partial_{n}\equiv(\vec{n}\cdot\vec{\nabla})$
is the derivate normal to the boundary.

\subsection{Analytical solution}

\subsubsection{Magnetic field of a circular loop in spherical harmonics}

% Fig. 1
\begin{figure}
\centering
\includegraphics[width=\columnwidth]{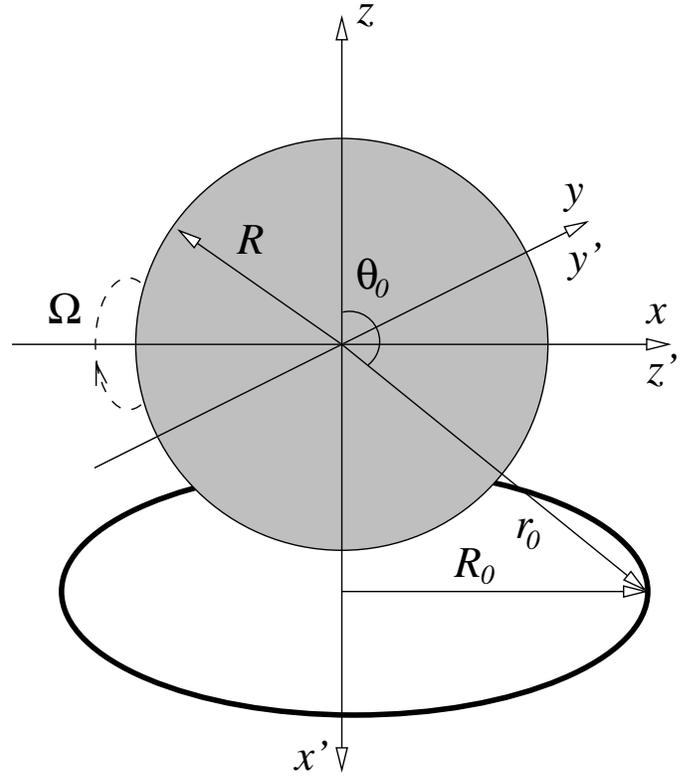}
% {sketch.eps}

\caption{\label{fig: sketch}Sketch to the formulation of the problem.}
\end{figure}

The magnetic field of an inductor consisting of a set of coaxial
circular loops represents a superposition of the fields of the
separate loops. Therefore, we first consider a single loop lying
parallel to the \textit{$x-y$} plane and centered with respect to
the \textit{z}-axis of a Cartesian coordinate system with a sphere
at its origin, see Fig. \ref{fig: sketch}. Approximation of a toroidal 
inductor of small cross-section by a circular loop is considered in 
more detail in Appendix 1. The loop is supplied
with an AC current with amplitude $I_{0}$ alternating harmonically
with a circular frequency $\omega$ as $I_{0}\cos(\omega t)$.
Henceforth we use the magnetic diffusion time
$\tau_{m}=\mu_{0}\sigma R^{2}$ and the radius of the sphere $R$ as
time and length scales while the current density and the vector
potential are scaled by $I_{0}/R^{2}$ and $\mu_{0}I_{0}$,
respectively. In the following, we assume all quantities to be
dimensionless with the same notation as for the dimensional
counterparts used so far. Further we consider the vector potential
generated by the loop
$\vec{A}^{e}(\vec{r},t)=\vec{A}_{0}^{e}(\vec{r})\cos(\bar{\omega}t)$
with the axisymmetric amplitude $\vec{A}_{0}^{e}(\vec{r}),$ where
$\bar{\omega}=\omega\mu_{0}\sigma R^{2}$ is the dimensionless AC
frequency which will be the main parameter throughout this paper.

The amplitude of the current density of such a loop can be represented
by using the Dirac $\delta$ function: 
\[
\vec{j}_{0}^{e}(\vec{r})=\vec{e}_{\phi}\delta(\theta-\theta_{0})\delta(r-r_{0})/r_{0}
\]
where $\theta_{0}$ and $r_{0}$ are the meridional angle and the
spherical radius of the loop with respect to the center of the sphere
as shown in Fig. \ref{fig: sketch}. Further, it is advantageous to
introduce a complex unit vector 
$\vec{e}_{\eta}=\frac{1}{\sqrt{2}}\left(\vec{e}_{x}+i\vec{e}_{y}\right)$
that allows us to represent the azimuthal unit vector as 
\[
\vec{e}_{\phi}=
\frac{i}{\sqrt{2}}\left(\vec{e}_{\eta}^{*}\e^{i\phi}-\vec{e}_{\eta}
\e^{-i\phi}\right),
\]
where $\vec{e}_{x}$ and $\vec{e}_{y}$ are $x$ and $y$ unit vectors,
respectively, and the asterisk denotes the complex conjugate. Then
the current density can be expanded in spherical harmonics as
\setlength{\arraycolsep}{0.0em}
\begin{eqnarray}
\nonumber
\vec{j}_{0}^{e}(\vec{r})&{}={}&-\sqrt{2}\pi\sin\theta_{0}\delta(r-r_{0})/r_{0}\\
&&{\times}\sum_{l=1}^{\infty}Y_{l1}(\theta_{0},0)\sum_{m=-1}^{1}\vec{I}^{m}Y_{lm}(\theta,\phi),
\label{eq:j0-sph}
\end{eqnarray}
\setlength{\arraycolsep}{5pt}%
where
$Y_{lm}(\theta,\phi)$ are spherical harmonics as defined in \cite{Jackson-75}. 
The expansion coefficients are 
\[
\vec{I}^{m}=\left\{ \begin{array}{ll} 0; & m\neq\pm1\\
i\vec{e}_{\eta}^{*}; & m=1\\
i\vec{e}_{\eta}; & m=-1\end{array}\right..
\]
Using the generating function of the spherical harmonics \cite{Jackson-75}
the vector potential of a circular loop is obtained as
\setlength{\arraycolsep}{0.0em}
\begin{eqnarray}
\nonumber
\vec{A}_{0}^{e}(\vec{r})&{}={}&\sqrt{2}\pi R_{0}
\sum_{l=1}^{\infty}\frac{Y_{l1}(\theta_{0},0)}{2l+1}
\frac{(r,r_{0})_{<}^{l}}{(r,r_{0})_{>}^{l+1}}\\
 && {\times}\sum_{m=-1}^{1}\vec{I}^{m}Y_{lm}(\theta,\phi),
\label{eq:A0e}
\end{eqnarray}
\setlength{\arraycolsep}{5pt}%
where $R_{0}=r_{0}\sin\theta_{0}$ is the cylindrical radius of the
loop; $(r,r')_{<}=\min(\left|\vec{r}\right|,\left|\vec{r}'\right|)$;
$(r,r')_{>}=\max(\left|\vec{r}\right|,\left|\vec{r}'\right|)$. In order to facilitate 
the following algebra, it is advantageous to introduce the functions
$X_{l}^{m}(\vec{r})=r^{-l-1}Y_{lm}(\theta,\phi)$ and
$\bar{X}_{l}^{m}(\vec{r})=r^{l}Y_{lm}(\theta,\phi)$, which are the
outer and the inner solutions of the Laplace equation associated
to the spherical harmonic $Y_{lm}(\theta,\phi)$. Basic properties
of those functions are given in Appendix 2. For the region occupied 
by the sphere ($r<r_0$) the solution above takes the form
\[
\vec{A}_{0}^{e}(\vec{r})=
\sqrt{2}\pi R_{0}\sum_{l=1}^{\infty}\frac{X_{l}^{1}(\vec{r}_{0})}{2l+1}
\sum_{m=-1}^{1}\vec{I}^{m}\bar{X}_{l}^{m}(\vec{r})
\]
where $\phi_{0}=0$ for $\vec{r}_{0}$ is assumed.

\subsubsection{Axisymmetric basic state}

For a sphere at rest the vector potential is sought as 
$\vec{A}(\vec{r},t)=\Re\left[\vec{A}_{0}(\vec{r})\e^{i\omega t}\right],$
where $\vec{A}_{0}$ is complex. For the interior of the sphere,
(\ref{eq:vecpot}) takes the form 
\begin{equation}
\nabla^{2}\vec{A}_{0}=i\bar{\omega}\vec{A}_{0}\label{eq:A0-ex}
\end{equation}
while for the exterior we have 
\begin{equation}
\nabla^{2}\vec{A}_{0}=0\label{eq:A0-in}
\end{equation}
whereas (\ref{eq:gauge}) takes the form of the Coulomb gauge
$\vec{\nabla}\cdot\vec{A}_{0}=0$. For the interior, the solution
is
\setlength{\arraycolsep}{0.0em}
\begin{eqnarray}
\nonumber
\vec{A}_{0}(\vec{r})&{}={}&\sqrt{2}\pi R_{0}\sum_{l=1}^{\infty}
\frac{X_{l}^{1}(\vec{r}_{0})}{2l+1}\bar{g}_{l}j_{l}(r\sqrt{\bar{\omega}/i})\\
 && {\times}\sum_{m=-1}^{1}\vec{I}^{m}Y_{lm}(\theta,\phi)
\label{eq:A0-insol}
\end{eqnarray}
\setlength{\arraycolsep}{5pt}%
where $\bar{g}_{l}$ are unknown coefficients to be determined from
the boundary conditions; $j_{l}(x)$ is the spherical Bessel function
of index $l$ \cite{AbramowitzStegun-65}. For the exterior, the solution
is represented as a superposition of external and induced fields 
$\vec{A}_{0}(\vec{r})=\vec{A}_{0}^{e}(\vec{r})+\vec{A}_{0}^{i}(\vec{r})$
where the latter is sought as 
\setlength{\arraycolsep}{0.0em}
\begin{equation}
\vec{A}_{0}^{i}(\vec{r}) = \sqrt{2}\pi R_{0}\sum_{l=1}^{\infty}
\frac{X_{l}^{1}(\vec{r}_{0})}{2l+1}
g_{l}\sum_{m=-1}^{1}\vec{I}^{m}X_{l}^{m}(\vec{r}).
\label{eq:A0-exsol}
\end{equation}
From the conditions (\ref{eq:bcnd-1}) applied at the surface of
the sphere at $r=1$ we find 
\[
\bar{g}_{l}=\frac{2l+1}{\sqrt{\bar{\omega}/i}}
\frac{1}{j_{l-1}(\sqrt{\bar{\omega}/i})};\qquad 
g_{l}=\frac{j_{l+1}(\sqrt{\bar{\omega}/i})}{j_{l-1}(\sqrt{\bar{\omega}/i})}.
\]
Some useful properties and an efficient algorithm for the calculation
of $g_{l}(x)$ and its derivative are given in Appendix 3.

The time-averaged total force on the sphere is 
\[
\vec{F}_{0}=\int_{V}\left\langle \vec{j}\times\vec{B}\right\rangle dV=
\frac{1}{2}\int_{V}\Re\left[\vec{j}_{0}\times\vec{B}_{0}^{*}\right]dV,
\]
where the integral is taken over the volume of the sphere $V$. Taking
into account that $\vec{j}_{0}=\vec{\nabla}\times\vec{B}_{0}^{i}$
and $\vec{B}_{0}=\vec{B}_{0}^{e}+\vec{B}_{0}^{i},$ where $\vec{B}_{0}^{e}$
is a purely real quantity, we obtain 
$\vec{F}_{0}=\frac{1}{2}\int_{V}\Re\left[\vec{j}_{0}\right]\times\vec{B}_{0}^{e}dV.$
According to the momentum conservation law the force exerted by a magnetic field 
on a body is opposite to that exerted by the body on the source of 
the magnetic field
$\vec{F}_{0}=-\frac{1}{2}\int_{\bar{V}}\vec{j}_{0}^{e}\times
\Re\left[\vec{B}_{0}^{i}\right]dV$
where the integral is taken over the space outside the sphere $\bar{V}.$
The last integral can be taken straightforwardly because the external
current (\ref{eq:j0-sph}) is defined in terms of a Dirac $\delta$
function. Taking into account that the induced magnetic field outside
the sphere is 
\setlength{\arraycolsep}{0.0em}
\begin{eqnarray*}
\vec{B}_{0}^{i} &{}={}& \sqrt{2}\pi R_{0}\sum_{l=2}^{\infty}
\frac{X_{l-1}^{1}(\vec{r}_{0})}{2l-1}g_{l-1}\sum_{m=-2}^{2}N_{l}^{m}X_{l}^{m}\\
&& {\times}\left(\frac{\vec{e}_{\eta}^{*}\times\vec{I}^{m-1}}{\sqrt{2}N_{l-1}^{m-1}}
-\frac{\vec{e}_{\eta}\times\vec{I}^{m+1}}{\sqrt{2}N_{l-1}^{m+1}}-
\frac{\vec{e}_{z}\times\vec{I}^{m}}{N_{l-1}^{m}}\right)
\end{eqnarray*}
\setlength{\arraycolsep}{5pt}%
where $N_{l}^{m}=\sqrt{\frac{(l-m)!(l+m)!}{2l+1}}$, we find 
\begin{equation}
\vec{F}_{0}=\vec{e}_{z}\pi\sum_{l=2}^{\infty}\Re\left[g_{l-1}\right]
f_{l-1}f_{l}\left(\frac{l^{2}-1}{4l^{2}-1}\right)^{1/2}
\label{eq:F0}
\end{equation}
where 
$f_{l}=r_{0}^{-l}\sin\theta_{0}\bar{P}_{l}^{1}(\cos\theta_{0})$
involves the normalised Legendre function $\bar{P}_{l}^{m}(x)$ 
related to the spherical harmonics as $Y_{lm}(\theta,\phi)=
\frac{(-1)^{m}}{\sqrt{2\pi}}\bar{P}_{l}^{m}(\cos\theta)e^{im\phi}$
\cite{AbramowitzStegun-65}. In case of $N$ current loops we have
\begin{equation}
f_{l}=\sum_{n=1}^{N}I_{n}r_{n}^{-l}\sin\theta_{n}\bar{P}_{l}^{1}(\cos\theta_{n})
\label{eq:fl}
\end{equation}
where $I_{n}$ is the dimensionless current carried by the \textit{n}-th
loop; $R_{n}$, $r_{n}$, and $\theta_{n}$ are defined analogously
as for the single loop. Note that the solution (\ref{eq:F0}) alone
could be obtained in a simpler way as in Ref. \cite{Smythe}. Our
following analysis, however, is aimed at non-axisymmetric solutions
which require a more complicated algebra.

\subsubsection{Perturbation due to a small displacement}

Consider a small displacement $\vec{x}$ of the center of the sphere
from the position $\vec{r}$ to $\vec{r}+\vec{x}.$ In the frame of
reference related to the sphere, this corresponds to a perturbation
of the external magnetic field 
\[
\vec{A}_{0}^{e}(\vec{r}+\vec{x}) \approx \vec{A}_{0}^{e}(\vec{r})+
(\vec{x}\cdot\vec{\nabla})\vec{A}_{0}^{e}(\vec{r})
=\vec{A}_{0}^{e}(\vec{r})+\left|\vec{x}\right|\vec{A}_{1}^{e}(\vec{r}),
\]
where 
\setlength{\arraycolsep}{0.0em}
\begin{eqnarray*}
\vec{A}_{1}^{e}(\vec{r})&{}={}&(\vec{\epsilon}_{x}\cdot
\vec{\nabla})\vec{A}_{0}^{e}(\vec{r})\\
&{}={}& \sqrt{2}\pi R_{0}\sum_{l=0}^{\infty}\frac{X_{l+1}^{1}(\vec{r}_{0})}{2l+1}
\sum_{m=-2}^{2}\vec{J}_{l}^{m}\bar{X}_{l}^{m}(\vec{r})
\end{eqnarray*}
\setlength{\arraycolsep}{5pt}%
$\vec{\epsilon}_{x}=\vec{x}/\left|\vec{x}\right|$ and
\setlength{\arraycolsep}{0.0em}
\begin{eqnarray*} 
\vec{J}_{l}^{m} &{}={}& \left(\vec{I}^{m-1}N_{l+1}^{m-1}
\frac{(\vec{\epsilon}_{x}\cdot\vec{e}_{\eta}^{*})}{\sqrt{2}}
+\vec{I}^{m}N_{l+1}^{m}(\vec{\epsilon}_{x}\cdot\vec{e}_{z})\right.\\
&&\left. {-}\:\vec{I}^{m+1}N_{l+1}^{m+1}\frac{(\vec{\epsilon}_{x}\cdot
\vec{e}_{\eta})}{\sqrt{2}}\right)/N_{l}^{m}.
\end{eqnarray*}
\setlength{\arraycolsep}{5pt}%
Now the external field represents a superposition of the axisymmetric
base field $\vec{A}_{0}^{e}(\vec{r})$ and a small but in general
non-axisymmetric perturbation $\vec{A}_{1}^{e}(\vec{r})$. The perturbation
of the inner field $\vec{A}_{1}$ is governed by the same Eqs. (\ref{eq:A0-ex},
\ref{eq:A0-in}) as the base field with the difference that now the
external field is given by $\vec{A}_{1}^{e}(\vec{r})$. Consequently,
the perturbation of the inner field and the corresponding induced
field outside the sphere can be obtained straightforwardly by replacing
the coefficients $\vec{I}^{m}$ by $\vec{J}_{l}^{m}$ in the corresponding
solutions (\ref{eq:A0-insol}, \ref{eq:A0-exsol}) for the base field.
Although the solution of the perturbed field obtained in such a way
satisfies (\ref{eq:A0-ex}, \ref{eq:A0-in})
and the boundary conditions, it turns out that the induced field outside
does not satisfy the Coulomb gauge unless the problem is axisymmetric,
\textit{i.e.}, the offset is along the symmetry axis. Thus, although
the Laplace equation is satisfied we formally have 
$\vec{\nabla}\times\vec{\nabla}\times\vec{A}_{1}^{i}=
\vec{\nabla}\vec{\nabla}\cdot\vec{A}_{1}^{i}=\vec{j}_{1}^{i}\neq0$
which implies that the current is leaking outside the sphere. In addition
note that the continuity of the tangential components of induction
following from the continuity of the vector potential and its normal
derivative at the surface of the sphere ensures continuity of the
radial, \textit{i.e.} normal, component of the current because 
$\vec{r}\cdot\vec{j}=\vec{r}\cdot\vec{\nabla}\times\vec{B}=
\vec{\nabla}\cdot(\vec{B}\times\vec{r})$.
Thus, the vector potential satisfying the Laplace equation and the
Coloumb gauge outside the sphere and being continuous at the surface
ensures that the current is closed in the sphere. As shown above,
the induced vector potential unambiguously follows from the imposed
one. Thus, the induced vector potential can be modified only by changing
the imposed one. This can be done by employing the gauge invariance
which allows us to add to the external vector potential $\vec{A}_{1}^{e}(\vec{r})$
a gradient of some gauge potential defined as 
\[
\Lambda_{1}^{e}(\vec{r})=\sqrt{2}\pi R_{0}\sum_{l=1}^{\infty}
\frac{X_{l}^{1}(\vec{r}_{0})}{2l+1}\sum_{m=-1}^{1}\lambda_{l}^{m}
\bar{X}_{l}^{m}(\vec{r}).
\]
In order to fulfill the Coulomb gauge for the external vector potential,
$\Lambda_{1}^{e}(\vec{r})$ has to be a harmonic function which is
reflected in the expression above. So, we obtain a set of free coefficients
$\lambda_{l}^{m}$ which can be chosen to satisfy the Coulomb gauge
for the induced field. This gauge transformation results in the replacement
of the original coefficients $\vec{J}_{l}^{m}$ by 
\setlength{\arraycolsep}{0.0em}
\begin{eqnarray*}
\vec{\mathcal{J}}_{l}^{m} &{}={}& \left(\vec{J}_{l}^{m}+
\frac{\vec{e}_{\eta}^{*}}{\sqrt{2}}N_{l+1}^{m-1}\lambda_{l+1}^{m-1}\right.\\
 &&\left. {-}\:\frac{\vec{e}_{\eta}}{\sqrt{2}}N_{l+1}^{m+1}\lambda_{l+1}^{m+1}+
\vec{e}_{z}N_{l+1}^{m}\lambda_{l+1}^{m}\right)/N_{l}^{m}.
\end{eqnarray*}
\setlength{\arraycolsep}{5pt}%
Then, similarly to the base field, the perturbation of the induced
field is obtained as 
\[
\vec{A}_{1}^{i}(\vec{r}) = \sqrt{2}\pi R_{0}\sum_{l=0}^{\infty}
\frac{X_{l-1}^{1}(\vec{r}_{0})}{2l-1}
g_{l}\sum_{m=-2}^{2}\vec{\mathcal{J}}_{l}^{m}\bar{X}_{l}^{m}(\vec{r}).
\]
From the Coulomb gauge we find 
\[
\lambda_{l}^{m}=(\vec{\epsilon}_{x}\cdot\vec{I}^{m})\frac{2l-1}{(2l+1)l}.
\]
The corresponding induction is 
\setlength{\arraycolsep}{0.0em}
\begin{eqnarray*}
\vec{B}_{1}^{i}(\vec{r}) &{}={}& \sqrt{2}\pi R_{0}\sum_{l=2}^{\infty}
\frac{X_{l}^{1}(\vec{r}_{0})}{2l-1}g_{l-1} 
\sum_{m=-3}^{3}N_{l}^{m}X_{l}^{m}(\vec{r})\\
 && {\times}\left[\frac{\vec{e}_{\eta}^{*}\times
\vec{\mathcal{J}}_{l-1}^{m-1}}{\sqrt{2}N_{l-1}^{m-1}}
-\frac{\vec{e}_{\eta}\times\vec{\mathcal{J}}_{l-1}^{m+1}}{\sqrt{2}N_{l-1}^{m+1}}
+\frac{\vec{e}_{z}\times\vec{\mathcal{J}}_{l-1}^{m}}{N_{l-1}^{m}}\right].
\end{eqnarray*}
\setlength{\arraycolsep}{5pt}%

According to our previous theory \cite{PriedeGerbeth-00o}, small-amplitude
oscillations of the sphere with $\vec{x}(t)=\vec{x}_{0}\cos(\Omega t)$
give rise to a perturbation of the force on the sphere 
\[
\vec{F}_{1}=\Re\left[(-\vec{K}
+i\Omega\vec{\G})\cdot\vec{x}_{0}\e^{i\Omega t}\right]
=-\vec{K}\cdot\vec{x}+\vec{\G}\cdot\frac{d\vec{x}}{dt}.
\]
where $\vec{K}$ and $-\vec{\G}$ are the effective electromagnetic
stiffness and damping tensors, respectively. Further we refer to the
elements of $\vec{K}$ and $\vec{\G}$ as the spring constants and
growth rates, respectively. Note that due to the axial symmetry both
$\vec{K}$ and $\vec{\G}$ are purely diagonal matrices with zero
azimuthal components. At first, the axial symmetry implies that a
purely axial displacement gives rise to a similar reaction force.
Second, if the displacement is purely radial the reaction force, on
the one hand, has to change to the opposite together with the displacement
because of linearity but, on the other hand, it has to rotate around
the symmetry axis together with the latter. Thus, a purely radial
displacement causes a purely radial reaction force and, therefore,
both $\vec{K}$ and $\vec{\G}$ are purely diagonal. Both diagonal
elements of $\vec{K}$ have to be positive for the static stability
whereas the corresponding growth rates have to be negative for dynamic
stability. Note that the growth rates can become positive and, thus,
destabilizing when the dimensionless AC frequency exceeds a certain
threshold depending on the configuration of the field \cite{PriedeGerbeth-00o}.
Both the spring constants and the growth rates can be calculated for
the given direction of the displacement specified by 
$\vec{\epsilon}_{x}={\vec{x}}/{\left|\vec{x}\right|}$
by using the corresponding base and perturbation fields \cite{PriedeGerbeth-00o}:
\begin{eqnarray*}
\vec{K}\cdot\vec{\epsilon}_{x} & = & -\frac{1}{2}\int_{V}
\left[\Re\left[\vec{j}_{1}\right]\times\vec{B}_{0}^{e}
+\Re\left[\vec{j}_{0}\right]\times\vec{B}_{1}^{e}\right]dV\\
\vec{\G}\cdot\vec{\epsilon}_{x} & = & \frac{1}{2}\partial_{\bar{\omega}}
\int_{V}\Im\left[\vec{j}_{1}\right]\times\vec{B}_{0}^{e}dV.
\end{eqnarray*}
Similarly to the integral force, it is advantageous to change the
region of integration from the sphere to the current loop 
\begin{eqnarray*}
\vec{K}\cdot\vec{\epsilon}_{x} & = & \frac{1}{2}\int_{\bar{V}}\left[\vec{j}_{0}^{e}
\times\Re\left[\vec{B}_{1}^{i}\right]+\vec{j}_{1}^{e}\times
\Re\left[\vec{B}_{0}^{i}\right]\right]dV,\\
\vec{\G}\cdot\vec{\epsilon}_{x} & = & -\frac{1}{2}\partial_{\bar{\omega}}
\int_{\bar{V}}\vec{j}_{0}^{e}\times\Im\left[\vec{B}_{1}^{i}\right]dV.
\end{eqnarray*}
After some algebra we find 
\setlength{\arraycolsep}{0.0em}
\begin{eqnarray}
\vec{\G}\cdot\vec{\epsilon}_{x} &{}={}& \pi\sum_{l=2}^{\infty}
\Im\left[\partial_{\bar{\omega}}g_{l-1}\right]\frac{f_{l}^{2}}{2l+1}(l^{2}-1)\\
 && {\times}\left[\frac{1}{2}\vec{e}_{z}\times\vec{\epsilon}_{x}\times
 \vec{e}_{z}(1-\frac{1}{l})+\vec{e}_{z}(\vec{\epsilon}_{x}\cdot\vec{e}_{z})\right].
\nonumber
\label{eq:Gamma}
\end{eqnarray}
\setlength{\arraycolsep}{5pt}%
Taking into account that 
\[\frac{1}{2}\int_{\bar{V}}\vec{j}_{1}^{e}\times
\vec{B}_{0}^{i}dV=-\frac{1}{2}\int_{\bar{V}}\vec{j}_{0}^{e}\times
(\vec{\epsilon}_{x}\cdot\vec{\nabla})\vec{B}_{0}^{i}dV
\]
because $\vec{j}_{1}^{e}(\vec{r})=(\vec{\epsilon}_{x}\cdot\vec{\nabla})
\vec{j}_{0}^{e}(\vec{r}),$
where $\vec{j}_{0}^{e}(\vec{r})$ is defined via Dirac $\delta$ functions
we eventually obtain 
\setlength{\arraycolsep}{0.0em}
\begin{eqnarray*}
\vec{K}\cdot\vec{\epsilon}_{x} &{}={}& -\pi\sum_{l=2}^{\infty}
\Re\left[g_{l-1}\right]\left\{\frac{f_{l}^{2}}{2l+1}(l^{2}-1)\right.\\
 &&\left. {\times}\left[\frac{1}{2}\vec{e}_{z}\times\vec{\epsilon}_{x}\times
\vec{e}_{z}(1-\frac{1}{l})+\vec{e}_{z}(\vec{\epsilon}_{x}\cdot
\vec{e}_{z})\right]\right.\\
 &&\left. {-}\:\frac{f_{l-1}f_{l+1}}{2l-1}\left(
\frac{l(l+2)(l^{2}-1)}{(2l-1)(2l+3)}\right)^{1/2}\right.\\
 &&\left. {\times}\left[\frac{1}{2}\vec{e}_{z}\times\vec{\epsilon}_{x}\times\vec{e}_{z}
-\vec{e}_{z}(\vec{\epsilon}_{x}\cdot\vec{e}_{z})\right]\right\}.
\end{eqnarray*}
\setlength{\arraycolsep}{5pt}%
Now, we can use this series to calculate both the stiffness and the
damping coefficients for a given coil defined by the coefficients
$f_{l}$ according to (\ref{eq:fl}).

\section{Spin-up instability}

\subsection{Governing equations}

We consider a solid sphere, as in Section 2, rotating with angular
velocity $\vec{\Omega}$ in an alternating magnetic field. The density
of the induced current is given by Ohm's law for a moving medium 
\[
\vec{j}=\sigma(\vec{E}+\vec{v}\times\vec{B})=\sigma(-\partial_{t}\vec{A}-
\vec{\nabla}\Phi+\vec{v}\times\vec{\nabla}\times\vec{A})
\]
where $\vec{v}=\vec{\Omega}\times\vec{r}$ is the velocity of a solid-body
rotation. The electric field induced by such a rotation may be represented as 
$\vec{v}\times\vec{\nabla}\times\vec{A}=\vec{\nabla}(\vec{v}\cdot\vec{A})-
(\vec{v}\cdot\vec{\nabla})\vec{A}-\vec{A}\times\vec{\Omega}.$
Then from the second Maxwell equation with neglected displacement
current, that corresponds to the quasistationary approximation assumed
throughout this study, we obtain 
\[
\partial_{t}\vec{A}+(\vec{v}\cdot\vec{\nabla})\vec{A}+\vec{A}\times
\vec{\Omega}=\frac{1}{\mu_{0}\sigma}\nabla^{2}\vec{A}.
\]
Note that this equation governs the vector potential in the laboratory
frame of reference where the body rotates while the source of the
magnetic field is at rest. Further we proceed to the frame of reference
rotating together with the sphere. In a rotating frame of reference,
where the sphere is at rest while the source of the magnetic field
rotates with velocity $-\vec{v}$, the equation above takes the form
as for a body at rest 
\[
\partial_{t}\vec{A}(\vec{r}',t)=\frac{1}{\mu_{0}\sigma}\nabla^{2}\vec{A}(\vec{r}',t),
\]
where $\vec{r}'$ is a radius vector, which is time-dependent in
the rotating frame of reference with $\partial_{t}\vec{r}'=\vec{\Omega}\times\vec{r}'$.
In a rotating frame of reference, the body rotation appears as an
additional time-dependence of the external magnetic field modulating
the time-dependence of the applied magnetic field.

The time-averaged total torque on the sphere is 
\[
\vec{M}=\frac{1}{2}\int_{V}\vec{r}\times\Re
\left[\vec{j}_{0}\right]\times\vec{B}_{0}^{e}dV.
\]
According to the conservation of the angular momentum the torque
exerted by the magnetic field on a body is opposite to that
exerted by the body on the source of the magnetic field
\[
\vec{M}=-\frac{1}{2}\int_{\bar{V}}\vec{r}\times\vec{j}_{0}^{e}\times
\Re\left[\vec{B}_{0}^{i}\right]dV,
\]
where the integral is taken over the volume $\bar{V}$ outside the
sphere. The last integral can be taken straightforwardly because 
$\vec{j}_{0}^{e}(\vec{r}')$ is defined via Dirac $\delta$ functions. 
Because $(\vec{r}\cdot\vec{j}_{0}^{e})=0$ the integral above may be 
represented as 
\[\vec{M}=-\frac{1}{2}\int_{\bar{V}}\vec{j}_{0}^{e}\Re
\left[\vec{r}\cdot\vec{\nabla}\times\vec{A}_{0}^{i}\right]dV
\]

\subsection{Analytical solution}

Further we consider a sphere rotating about an axis perpendicular
to the symmetry axis of an axisymmetric external magnetic field of
a circular loop. The coordinate system related to the loop with the
\textit{$z$} axis as symmetry axis is chosen so that the rotation
takes place about the $x$-axis, as shown in Fig. \ref{fig: sketch}.
It is convenient to analyze this rotation in a coordinate system turned
about the $y$-axis by 90 degrees so that the \textit{$z'$}-axis
of the transformed coordinate system coincides with the original $x$-axis
which is the axis of rotation and, respectively, the \textit{$x'$}-axis
is directed opposite to the \textit{$z$} axis of the original coordinate
system. Then the basis vectors of both coordinate systems are related
as $\vec{e}_{x}'=-\vec{e}_{z},$ $\vec{e}_{y}'=\vec{e}_{y},$ and
$\vec{e}_{z}'=\vec{e}_{x}.$ Respectively, the complex unity vector is 
$\vec{e}_{\eta}=\frac{1}{2}\left(\vec{e}_{\eta}'-\vec{e}_{\eta}'^{*}\right)+
\frac{\vec{e}_{z}'}{\sqrt{2}},$ where 
$\vec{e}_{\eta}'=\frac{1}{\sqrt{2}}\left(\vec{e}_{x}'+i\vec{e}_{y}'\right).$
To represent the vector potential of a circular loop in the turned
coordinate system we use the addition theorem for Legendre polynomials
\cite{AbramowitzStegun-65} written as 
\[
\bar{X}_{l}^{0}(\vec{r})=\sqrt{\frac{4\pi}{2l+1}}
\sum_{m=-l}^{l}Y_{lm}(\pi/2,0)\bar{X}_{l}^{m}(\vec{r}')
\]
together with the relation 
\[
\bar{X}_{l}^{1}(\vec{r})=\frac{2l+1}{2l+3}\frac{N_{l}^{1}}{N_{l+1}^{0}}
\sqrt{2}(\vec{e}_{\eta}\cdot\vec{\nabla})\bar{X}_{l+1}^{0}(\vec{r})
\]
which yields 
\setlength{\arraycolsep}{0.0em}
\begin{eqnarray*}
\vec{A}_{0}^{e}(\vec{r}') &{}={}& \sqrt{2}\pi R_{0} \sum_{l=1}^{\infty}
\frac{X_{l}^{1}(\vec{r}_{0})}{2l+1}\frac{N_{l}^{1}}{N_{l+1}^{0}}
\sqrt{\frac{4\pi}{2l+3}}\\
 && {\times}\sum_{m=-l}^{l} \vec{I}_{l}^{m}\bar{X}_{l}^{m}(\vec{r}')
\end{eqnarray*}
\setlength{\arraycolsep}{5pt}%
where 
\setlength{\arraycolsep}{0.0em}
\begin{eqnarray*}
\vec{I}_{l}^{m} &{}={}& \frac{1}{N_{l}^{m}}\left[
\left(\vec{e}_{\eta}'-\vec{e}_{\eta}'^{*}\right)N_{l+1}^{m}Y_{l+1,m}(\pi/2,0)\right.\\
 &&{+}\:i\frac{\vec{e}_{z}'}{\sqrt{2}}\left(N_{l+1}^{m-1}Y_{l+1,m-1}(\pi/2,0)\right.\\
 &&\left.\left. {-}\:N_{l+1}^{m+1}Y_{l+1,m+1}(\pi/2,0)\right)\right].
\end{eqnarray*}
\setlength{\arraycolsep}{5pt}%
Similarly, the current density of the loop can be rewritten as 
\setlength{\arraycolsep}{0.0em}
\begin{eqnarray*}
\vec{j}_{o}^{e}(\vec{r}') &{}={}& -\sqrt{2}\pi\sin\theta_{0}\frac{\delta(r-r_{0})}{r_{0}}\\
 && {\times}\sum_{l=1}^{\infty}Y_{l1}(\theta_{0},0)\frac{N_{l}^{1}}{N_{l+1}^{0}}
 \sqrt{\frac{4\pi}{2l+3}}\\
 && {\times}\sum_{m=-l}^{l}\vec{I}_{l}^{m}Y_{lm}(\theta',\phi').
\end{eqnarray*}
\setlength{\arraycolsep}{5pt}%
In this case, we define a gauge potential 
\[
\Lambda_{0}^{e}(\vec{r}')=\sqrt{2}\pi\sum_{l=2}^{\infty}
\frac{2l-1}{2l+1}\bar{f}_{l-1}\sum_{m=-l}^{l}\lambda_{l}^{m}\bar{X}_{l}^{m}(\vec{r}')
\]
where $\bar{f}_{l}=R_{0}\frac{X_{l}^{1}(\vec{r}_{0})}{2l+1}
\frac{N_{l}^{1}}{N_{l+1}^{0}}\sqrt{\frac{4\pi}{2l+1}};$
$\lambda_{l}^{m}$ are unknown coefficients to be determined subsequently
from the Coulomb gauge for the vector potential induced outside the
sphere. Addition of $\vec{\nabla}\Lambda_{0}^{e}$ to $\vec{A}_{0}^{e}$
results in the replacement of the coefficients $\vec{I}_{l}^{m}$
by 
\setlength{\arraycolsep}{0.0em}
\begin{eqnarray*}
\vec{\mathcal{I}}_{l}^{m} &{}={}& \vec{I}_{l}^{m}+\left(\frac{\vec{e}_{\eta}^{*}}
{\sqrt{2}}N_{l+1}^{m-1}\lambda_{l+1}^{m-1}
-\frac{\vec{e}_{\eta}}{\sqrt{2}}N_{l+1}^{m+1}\lambda_{l+1}^{m+1}\right.\\
 &&\left. {+}\:\vec{e}_{z}N_{l+1}^{m}\lambda_{l+1}^{m}\right)/N_{l}^{m}.
\end{eqnarray*}
\setlength{\arraycolsep}{5pt}%
Now we can simply proceed to the frame of reference rotating together
with the sphere by redefining the azimuthal angle as $\phi'=\phi''+\bar{\Omega}t,$
where $\phi''$ is the azimuthal angle in the rotating frame of reference
and $\bar{\Omega}$ is the dimensionless angular velocity of rotation.
Respectively, the complex unit vector and the spherical harmonics
change in a rotating frame of reference as 
$\vec{e}_{\eta}'=\vec{e}_{\eta}''\e^{i\bar{\Omega}t}$
and $\bar{X}_{l}^{m}(\vec{r}')=\bar{X}_{l}^{m}(\vec{r}'')\e^{im\bar{\Omega}t}.$
In a rotating frame of reference, the vector potential takes the form
\setlength{\arraycolsep}{0.0em}
\begin{eqnarray*}
\vec{A}_{0}^{e}(\vec{r}'') &{}={}& \sqrt{2}\pi\sum_{l=1}^{\infty}\bar{f}_{l}
\sum_{m=-l}^{l}\left(\vec{e}_{z}''(\vec{e}_{z}'\cdot
\vec{\mathcal{I}}_{l}^{m})\e^{im\bar{\Omega}t}\right.\\
 && {+}\:\vec{e}_{\eta}''(\vec{e}_{\eta}'^{*}\cdot
\vec{\mathcal{I}}_{l}^{m})\e^{i(m+1)\bar{\Omega}t}\\
 &&\left. {+}\:\vec{e}_{\eta}''^{*}(\vec{e}_{\eta}'\cdot
\vec{\mathcal{I}}_{l}^{m})\e^{i(m-1)\bar{\Omega}t}\right)\bar{X}_{l}^{m}(\vec{r}'').
\end{eqnarray*}
\setlength{\arraycolsep}{5pt}%
Further we consider the current alternating harmonically with the
dimensionless frequency $\bar{\omega}$ as $\vec{j}^{e}(\vec{r}',t)=
\vec{j}_{0}^{e}(\vec{r}')\cos(\bar{\omega}t)$
and the vector potential alternating in the laboratory frame of reference,
respectively, as 
\[
\vec{A}^{e}(\vec{r}',t)=\vec{A}_{0}^{e}(\vec{r}')\cos(\bar{\omega}t)=
\Re\left[\vec{A}_{0}^{e}(\vec{r}')\e^{i\bar{\omega}t}\right].
\]
In the laboratory frame of reference, the complex amplitude of the
vector potential induced outside the sphere is obtained as 
\setlength{\arraycolsep}{0.0em}
\begin{eqnarray*}
\vec{A}_{0}^{i}(\vec{r}')&{}={}&\sqrt{2}\pi\sum_{l=1}^{\infty}\bar{f}_{l}\sum_{m=-l}^{l}
\left(\vec{e}_{\eta}'(\vec{e}_{\eta}'^{*}\cdot\vec{\mathcal{I}}_{l}^{m})g_{l}^{m+1}\right.\\
 &&{+}\:\vec{e}_{\eta}'^{*}(\vec{e}_{\eta}'\cdot\vec{\mathcal{I}}_{l}^{m})g_{l}^{m-1}\\
 &&\left. {+}\:\vec{e}_{z}'(\vec{e}_{z}'\cdot\vec{\mathcal{I}}_{l}^{m})g_{l}^{m}\right)
X_{l}^{m}(\vec{r}')
\end{eqnarray*}
\setlength{\arraycolsep}{5pt}%
where 
$g_{l}^{m}=\frac{j_{l+1}(\sqrt{\bar{\omega}_{m}/i})}
{j_{l-1}(\sqrt{\bar{\omega}_{m}/i})}$
and $\bar{\omega}_{m}=\bar{\omega}+m\bar{\Omega}$. From the Coulomb
gauge $\vec{\nabla}\cdot\vec{A}_{0}^{i}=0$ we find 
\setlength{\arraycolsep}{0.0em}
\begin{eqnarray*}
\lambda_{l+1}^{m}N_{l+1}^{m}\left[\frac{1}{2\left(N_{l}^{m-1}\right)^{2}}
+\frac{1}{2\left(N_{l}^{m+1}\right)^{2}}+
\frac{1}{\left(N_{l}^{m}\right)^{2}}\right]&{}={}&\\ 
{}={}\left[
 \frac{(\vec{e}_{\eta}'^{*}\cdot\vec{I}_{l}^{m-1})}{\sqrt{2}\left(N_{l}^{m-1}\right)^{2}}
-\frac{(\vec{e}_{\eta}'\cdot\vec{I}_{l}^{m+1})}{\sqrt{2}\left(N_{l}^{m+1}\right)^{2}}
-\frac{(\vec{e}_{z}'\cdot\vec{I}_{l}^{m})}{\left(N_{l}^{m}\right)^{2}}
\right]&& 
\end{eqnarray*}
\setlength{\arraycolsep}{5pt}%
For the field induced outside the sphere we obtain 
\setlength{\arraycolsep}{0.0em}
\begin{eqnarray*}
(\vec{r}'\cdot\vec{\nabla}\times\vec{A}_{0}^{i}) &{}={}& i\pi\sum_{l=1}^{\infty}
l\bar{f}_{l}\sum_{m=-l}^{l}g_{l}^{m}X_{l}^{m}(\vec{r}')\\
 &&{\times}\left[N_{l+1}^{m-1}Y_{l+1}^{m-1}(\pi/2,0)\right.\\
 &&\left. {+}\:N_{l+1}^{m+1}Y_{l+1}^{m+1}(\pi/2,0)\right].
\end{eqnarray*}
\setlength{\arraycolsep}{5pt}%
The gauge potential $\Lambda_{0}^{e}$ vanishes in the expression
above and so it does not affect the total torque which is 
\setlength{\arraycolsep}{0.0em}
\begin{eqnarray*}
\vec{M} &{}={}& \vec{e}_{z}'\frac{\pi}{2}\sum_{l=1}^{\infty}
\frac{f_{l}^{2}}{(2l+1)^{2}(l+1)}\sum_{m=-l}^{l}\Im\left[g_{l}^{m}\right]\\
 && {\times}\left[(l-m+1)(l-m+2)\bar{P}_{l+1}^{m-1}(0)^{2}\right.\\ 
 &&\left. {-}\:(l+m+1)(l+m+2)\bar{P}_{l+1}^{m+1}(0)^{2}\right].
\end{eqnarray*}
\setlength{\arraycolsep}{5pt}%
Other components of the torque, which are not parallel to the angular
velocity of rotation, vanish because of $\bar{P}_{l}^{m}(0)\bar{P}_{l}^{m\pm1}(0)=0$.
For slow rotations with $\bar{\Omega}\ll\bar{\omega}$ we have 
\[
g_{l}^{m}-g_{l}^{-m}=2m\bar{\Omega}\partial_{\bar{\omega}}g_{l}+O(\bar{\Omega}^{3})
\]
that results in $\vec{M}\approx\vec{e}_{z}'\bar{\Omega}\partial_{\bar{\Omega}}M$,
where 
\setlength{\arraycolsep}{0.0em}
\begin{eqnarray}
\label{eq:dM}
\partial_{\bar{\Omega}}M &{}={}& \pi\sum_{l=1}^{\infty}\frac{\Im\left[
\partial_{\bar{\omega}}g_{l}\right]f_{l}^{2}}{(2l+1)^{2}(l+1)}\sum_{m=1}^{l}m\\
 && {\times}\left[(l-m+1)(l-m+2)\bar{P}_{l+1}^{m-1}(0)^{2}\right.\nonumber\\
 &&\left. {-}\:(l+m+1)(l+m+2)\bar{P}_{l+1}^{m+1}(0)^{2}\right].\nonumber
\end{eqnarray}
\setlength{\arraycolsep}{5pt}%
For the rest state of the sphere to be stable to rotational perturbations,
the torque caused by these perturbations has to damp them that corresponds
to $\partial_{\bar{\Omega}}M<0.$ In the opposite case, the rest state
is unstable. The state is marginally stable when 
$\partial_{\bar{\Omega}}M(\bar{\omega}_{c})=0,$
which is the equation defining the critical frequency $\bar{\omega}_{c}$
for the spin-up instability.

\section{Numerical results}

\subsection{Small-amplitude oscillations}

\begin{figure*}
\centering
\subfigure[]{\includegraphics[width=\columnwidth]{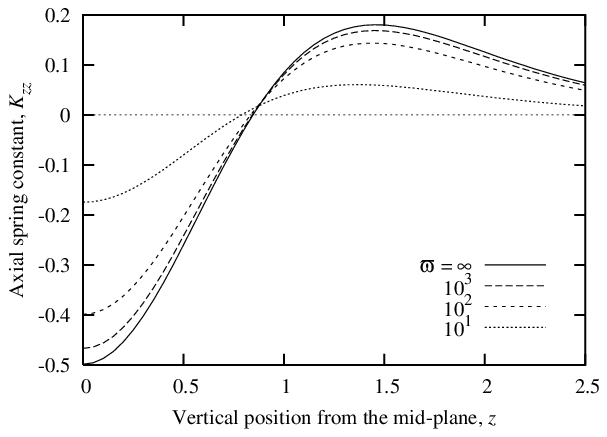}}
%{zspirngR2H0s.eps}}

\centering
\subfigure[]{\includegraphics[width=\columnwidth]{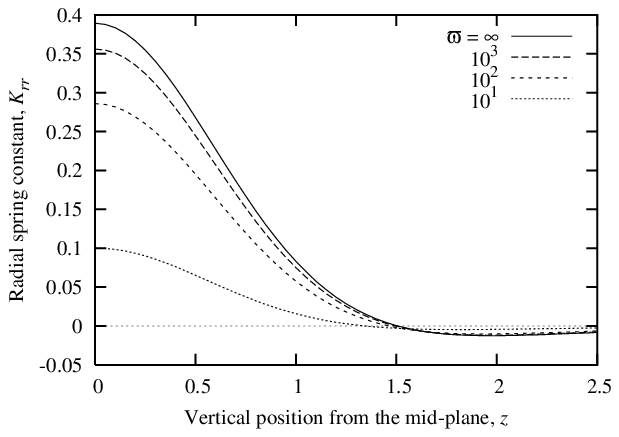}}%
%{rspirngR2H0s.eps}}
\subfigure[]{\includegraphics[width=\columnwidth,]{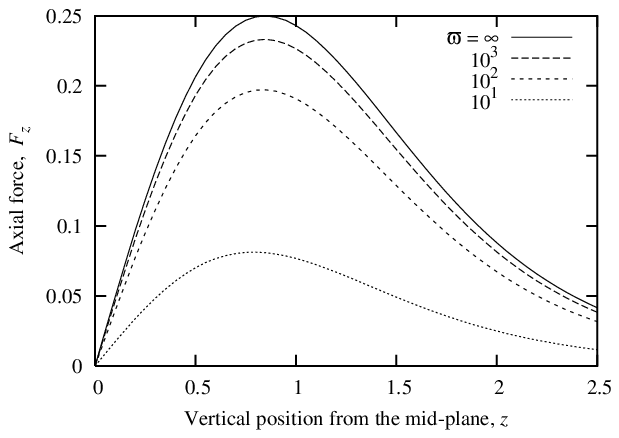}}
%{frcR2H0s.eps}}

\caption{\label{fig: force} Force (\textit{a}), axial (\textit{b}) and radial
(\textit{c}) spring constants in the magnetic field of a single loop
with $R=2$ versus the position of the sphere from the mid-plane at
various dimensionless frequencies. A negative spring constant implies
that the position is statically unstable.}
\end{figure*}

In the following we present some numerical results for an inductor
consisting of two circular loops of equal radii placed coaxially at
the distance $H$ from the mid-plane. The current having the same
amplitude may flow either in the same or in opposite directions in
both loops. Subsequently, both these cases are referred to as symmetric
and antisymmetric, respectively. We start with a simple symmetric
case when both loops coincide ($H=0$) making up a single loop with
doubled current amplitude. The forces along with axial and radial
spring constants versus the distance of the sphere from such a loop
of dimensionless radius $R=2$ are shown in Fig. \ref{fig: force}
for various dimensionless frequencies. As seen in Fig. \ref{fig: force}(\textit{a}),
at large enough distances the force, which is repulsing, raises with
decreasing distance between the loop and the sphere. At some distance,
the force attains a maximum and tends to zero as the sphere approaches
the center of the loop. Such a reducing force implies that the position
of the sphere is statically unstable to axial perturbations. The position
is statically stable to axial perturbations when the axial spring
constants shown in Fig. \ref{fig: force}(\textit{b}) are positive
that requires the sphere to be placed further away from the loop than
the point of maximum force which is at $z\approx0.75$. A position
being axially stable may become unstable radially when the distance
from the loop becomes too large (see Fig. \ref{fig: force}\textit{b}).
Thus the range of statically stable positions for a single loop is
rather limited ($0.75<z<1.5$) and does not change significantly with
the frequency of the magnetic field.

% Fig. 2

\begin{figure*}
\centering
\includegraphics[width=\columnwidth]{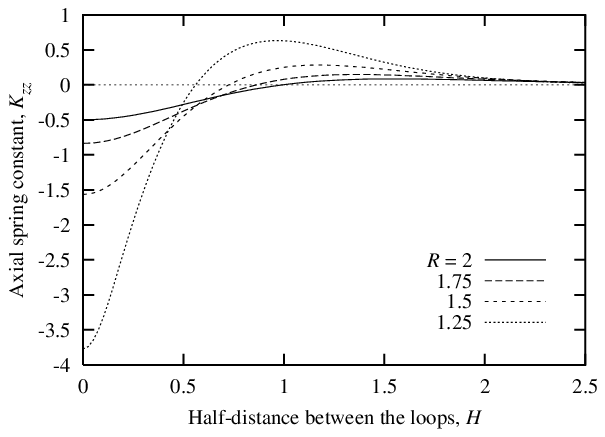}
%{zsprinfZ0s.eps}
\put (0,0){(a)}
\includegraphics[width=\columnwidth]{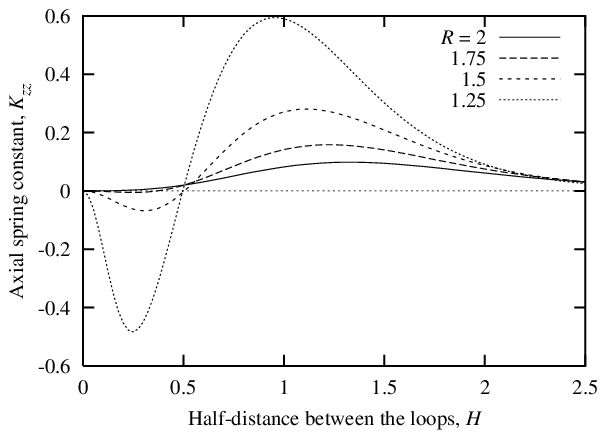}
%{zsprinfZ0a.eps}

\centering
\includegraphics[width=\columnwidth]{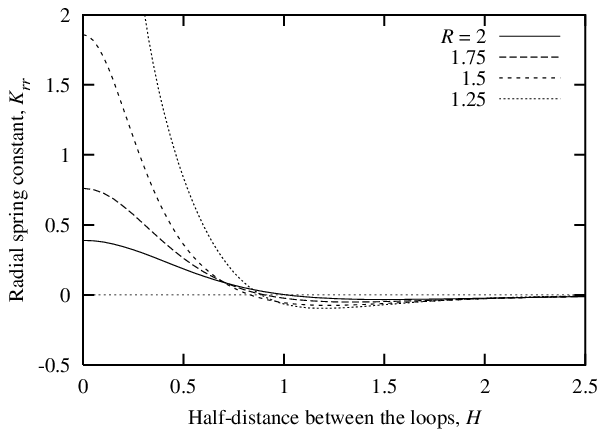}
%{rsprinfZ0s.eps}
\put(0,0){(b)}
\includegraphics[width=\columnwidth]{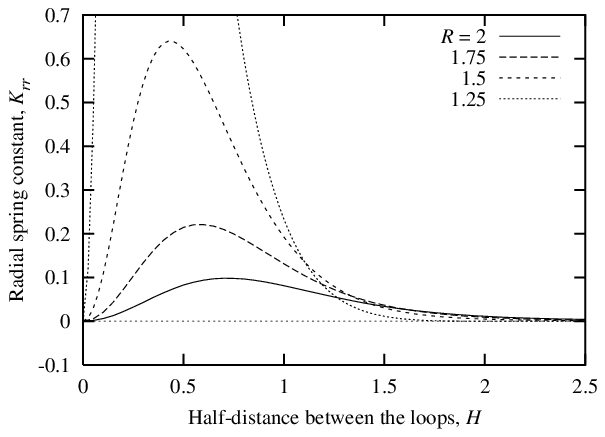}
%{rsprinfZ0a.eps}

\caption{\label{fig: finfZ0} Axial (a) and radial (b) spring constants versus
the half-distance between the loops for both symmetric (on left) and
antisymmetric (on right) arrangements at various radii of the loops
in perfect-conductor approximation ($\bar{\omega}=\infty$).}
\end{figure*}

Further let us turn briefly to the static stability of a sphere in
an arrangement of two coaxial current loops of the same radius placed
symmetrically at distance $H$ from the center of the sphere. For
the sake of brevity we restrict the static stability analysis to the
perfect-conductor approximation $\bar{\omega}=\infty.$ Axial and
radial spring constants are shown in Fig. \ref{fig: finfZ0} versus
the distance $H$ at various radii of the loops for both symmetric
and antisymmetric arrangements. In the symmetric arrangement, which
is usually used for heating, the position of the sphere is statically
unstable in radial direction when the loops are too much separated,
and it becomes unstable in the axial direction when the loops are
moved too tightly together. Overlapping of the stability ranges for
radial and axial perturbations depends on the radius of the loops.
For $R=2$ the range of axial stability ends at $H\approx1$ where
the range of radial stability begins. Thus, there is no overlapping
of the stability ranges in this case that implies a static instability
regardless of the distance between the loops. A range of positions
statically stable to both axial and radial perturbations is possible
only for sufficiently small radii of the loops. For instance at $R=1.25$
this range is approximately $0.5<z<0.75.$ In contrast to the symmetric
arrangement, an antisymmetric one, which is usually used for positioning
of the sample, ensures a statically stable state to radial perturbations
regardless of the distance between the loops (see Fig. \ref{fig: finfZ0}\textit{b}
on right). The position of the sphere is stable to axial perturbations
for any radius of the loop provided the loops are separated by a distance
larger than the radius of the sphere ($H>0.5$). There might be an
axial instability for smaller separations of the loops at $R<2$ (see
Fig. \ref{fig: finfZ0}\textit{a} on right).

% Fig. 3

% Fig. 4
\begin{figure*}
\centering
\includegraphics[width=\columnwidth]{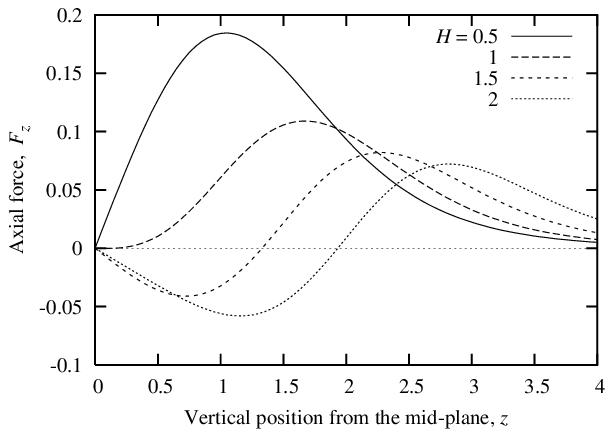}
%{finfR2s.eps}
{\put(0,0){(a)}}
\includegraphics[width=\columnwidth]{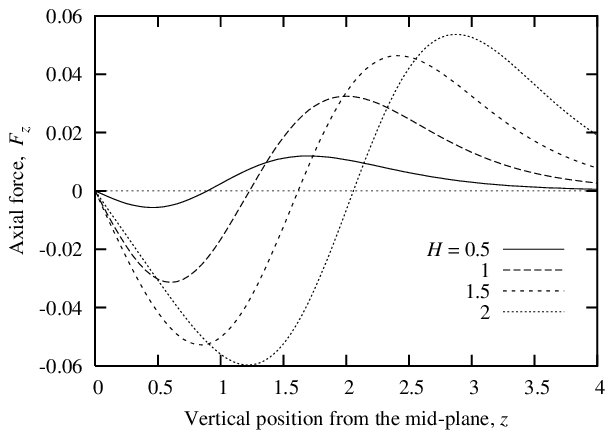}
%{finfR2a.eps}

\centering
\includegraphics[width=\columnwidth]{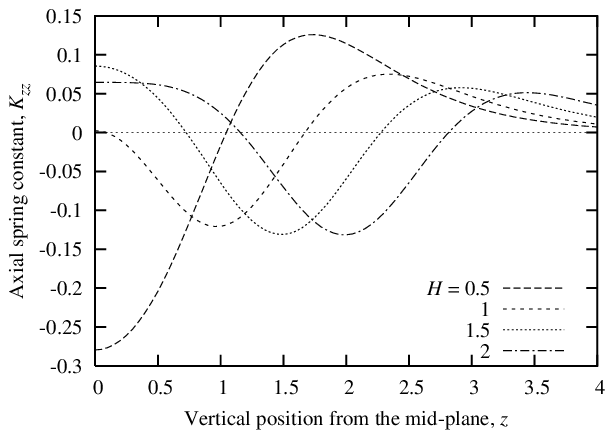}
%{zsprinfR2s.eps}
{\put(0,0){(b)}}
\includegraphics[width=\columnwidth]{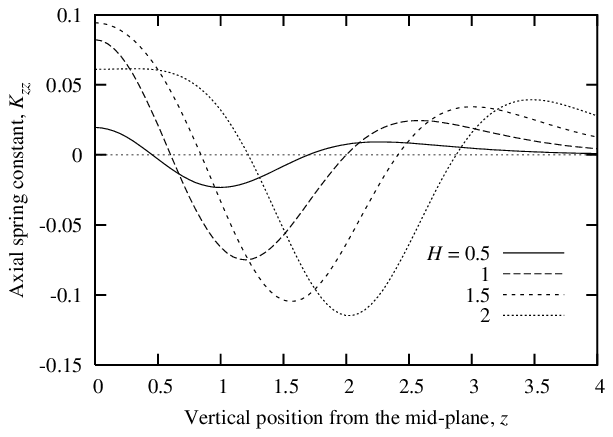}
%{zsprinfR2a.eps}

\centering
\includegraphics[width=\columnwidth]{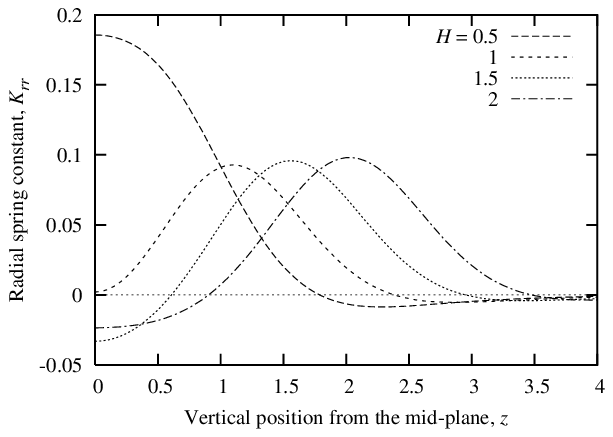}
%{rsprinfR2s.eps}
{\put(0,0){(c)}}
\includegraphics[width=\columnwidth]{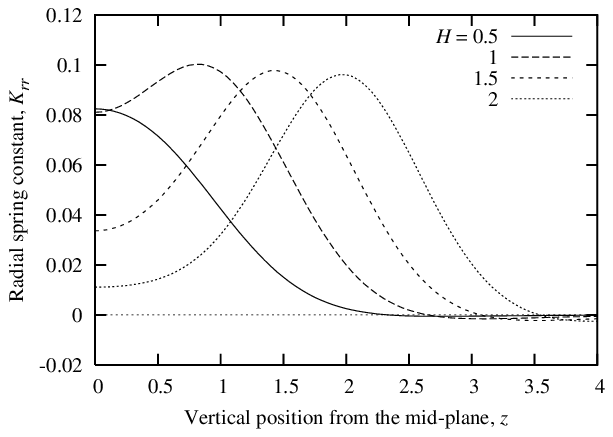}
%{rsprinfR2a.eps}

\caption{\label{fig: finfR2} Force (a), axial (b) and radial (c) spring constants
versus the position of the sphere above the mid-plane for symmetric
(on left) and antisymmetric (on right) arrangements of two coaxial
loops of radius $R=2$ and various distances from the mid-plane at
$\bar{\omega}=\infty$.}
\end{figure*}

Forces together with axial and radial spring constants versus the
distance of the sphere from the mid-plane of both symmetric and antisymmetric
arrangements of two loops are shown in Fig. \ref{fig: finfR2} for
various distances between the loops in the perfect-conductor approximation
($\bar{\omega}=\infty$). It is seen that the position of the sphere
in the middle-point ($z=0$) of a symmetric arrangement (Fig. \ref{fig: finfR2}
on left) is unstable either in radial or axial directions for almost
all distances between the coils. However, there is one distance between
the coils, $H\approx1$, at which both spring constants cross zero
almost simultaneously. Thus, this distance is an optimal one for the
static stability of the sphere in the middle-point. Note that the
distance $H=R/2$ provides the most uniform magnetic field in the
vicinity of the middle-point of the symmetric arrangement. This is
because the next to the leading order contribution for the magnetic
field of symmetric arrangement vanishes at the aspect ratio $H/R=\cot\theta_{0}=0.5$,
\textit{i.e}., $P_{3}^{1}(\cos\theta_{0})=0$ in (\ref{eq:A0e}).
Similarly, the distance $H=R\sqrt{3}/2$ providing the most uniform
gradient of the magnetic field in the vicinity of the middle-point
of an antisymmetric arrangement is expected to ensure the widest range
of statically stable positions for loops of large enough radius.

% Fig. 5
\begin{figure*}
\centering
\includegraphics[height=0.23\textheight]{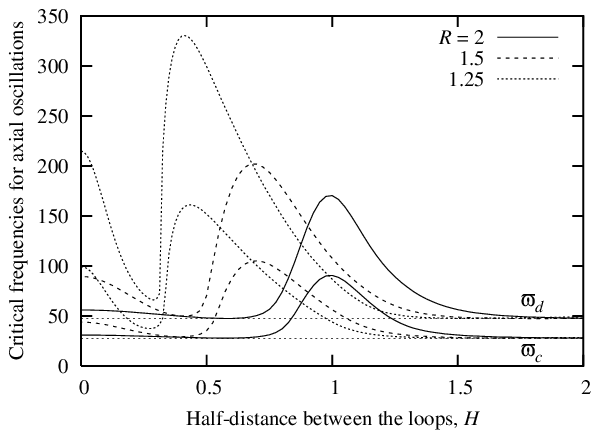}
%{osczZ0s.eps}
{\put(0,0){(a)}}
\includegraphics[height=0.23\textheight]{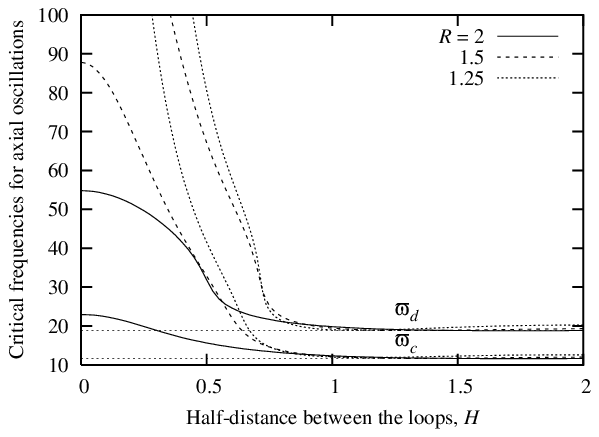}
%{osczZ0a.eps}

\centering
\includegraphics[height=0.23\textheight]{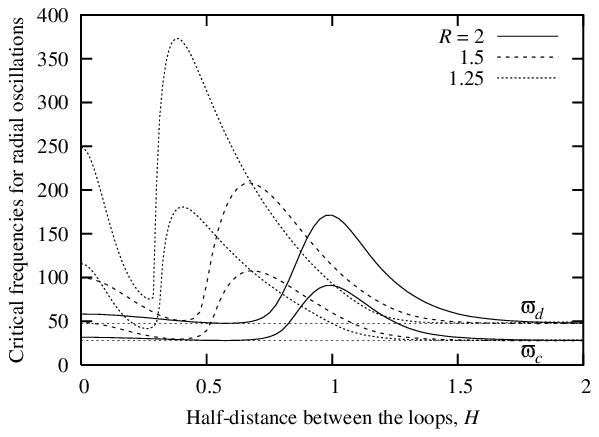}
%{oscrZ0s.eps}
{\put(0,0){(b)}}
\includegraphics[height=0.23\textheight]{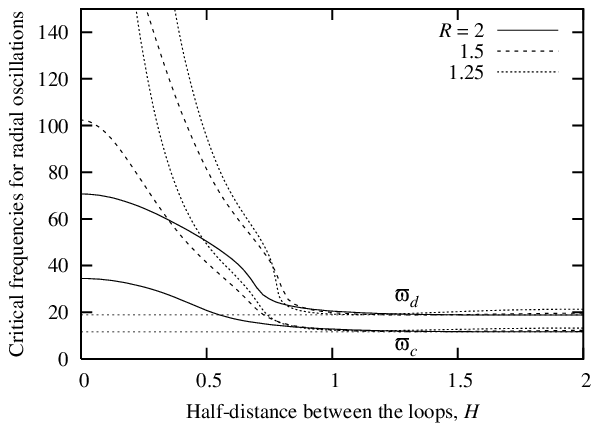}
%{oscrZ0a.eps}

\centering
\includegraphics[height=0.23\textheight]{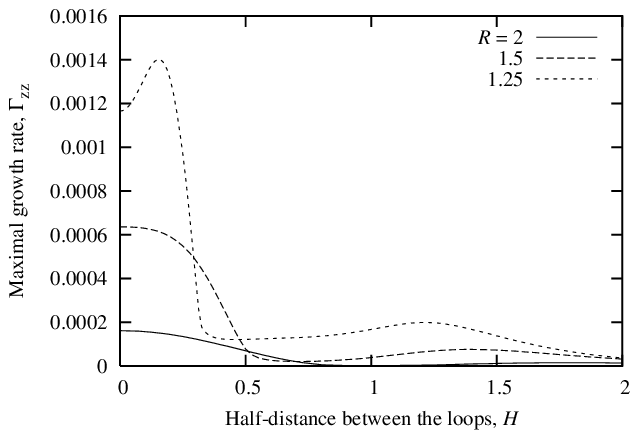}
%{osczmaxZ0s.eps}
{\put(0,0){(c)}}
\includegraphics[height=0.23\textheight]{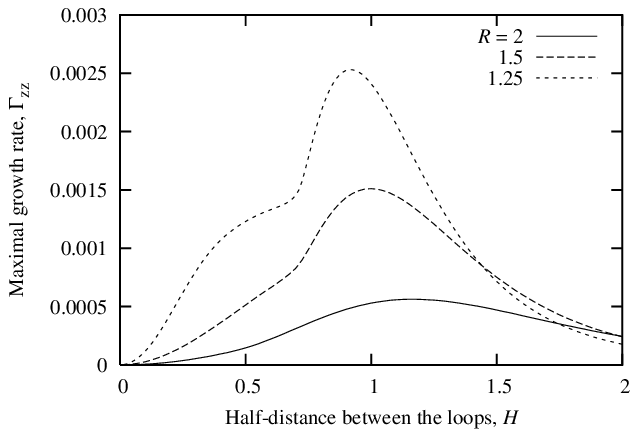}
%{osczmaxZ0a.eps}

\centering
\includegraphics[height=0.23\textheight]{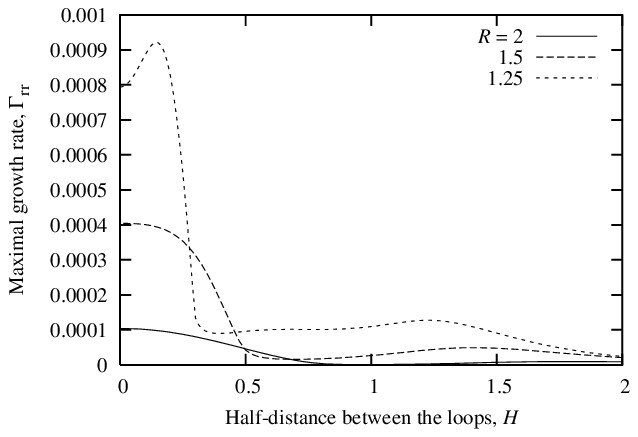}
%{oscrmaxZ0s.eps}
{\put(0,0){(d)}}
\includegraphics[height=0.23\textheight]{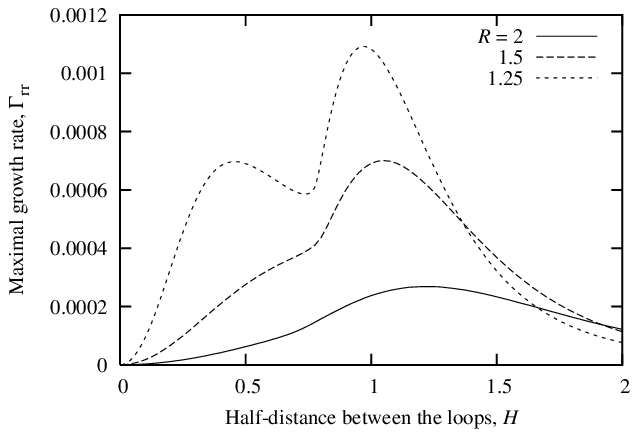}
%{oscrmaxZ0a.eps}

\caption{\label{fig: oscZ0} Critical ($\bar{\omega}_{c}$) and most dangerous
($\bar{\omega}_{d}$) dimensionless AC frequencies (\textit{a}, \textit{b})
and maximal growth rates (\textit{c}, \textit{d}) for axial (\textit{a},
\textit{c}) and radial (\textit{b}, \textit{d}) oscillations of a
sphere about the middle-point of symmetric (on left) and antisymmetric
(on right) arrangements of two loops versus the half-distance between
the loops at various radii of the loops. }
\end{figure*}

The dynamic stability of small-amplitude radial and axial oscillations
is characterized by the corresponding growth rates $\G_{rr}$ and
$\G_{zz}$ defined by (\ref{eq:Gamma}). The threshold of linear
stability is given by the frequency above which the corresponding
growth rate becomes positive. At the threshold frequency, the amplitude
of oscillations turns from decaying to growing in time. A characteristic
feature of this instability is that the growth rate, becoming positive
above the threshold frequency, reaches a maximum at a certain frequency
and tends to zero at higher frequencies. The frequency at which the
growth rate attains a maximum is referred to as the most dangerous
one. Threshold frequencies along with the most dangerous one and the
corresponding maximal growth rates are plotted in Fig. \ref{fig: oscZ0}
versus the half-distance between the loops at various radii of the
loops for both axial and radial oscillations of a sphere about the
middle-point of both symmetric and antisymmetric arrangements of two
loops. The most dangerous frequency is always above the threshold
one. As seen in Figs. \ref{fig: oscZ0}(\textit{a}) and \ref{fig: oscZ0}(\textit{b}),
at large enough distances between the loops, the threshold and the
most dangerous frequencies tend to $\bar{\omega}_{c}=27.682$ and
$\bar{\omega}_{d}=47.196$ in the symmetric arrangement, and to $\bar{\omega}_{c}=11.609$
and $\bar{\omega}_{d}=18.792$ in the antisymmetric one, respectively.
Note that these critical frequencies for the oscillatory instability
in spatially quadratic and linear magnetic fields coincide with those
for the spin-up instability in linear and uniform magnetic fields,
respectively, which will be considered later. For each radius of a
symmetric arrangement, there is a certain optimal distance between
the loops that maximizes the critical frequency. For large enough
radii the optimal distance between the loops is approximately equal
to the radius, $2H=R$, which is optimal also for the static stability
as considered above. With decreasing of the loop radius, the optimal
distance becomes slightly larger (see Figs. \ref{fig: oscZ0}\textit{a}
and \ref{fig: oscZ0}\textit{b} on left). In the antisymmetric case,
the critical frequency changes weakly on reducing of the distance
$H$ until the loops approach a distance $H\approx0.75$ to the mid-plane
(see Figs. \ref{fig: oscZ0}\textit{a} and \ref{fig: oscZ0}\textit{b}
on right). Further approaching of the loops results in the increase
of the critical frequency which is the larger, the smaller the radius.
This raise is caused by the nonuniformity of the magnetic field in
the vicinity of the loop. Although the critical frequency for axial
oscillations is, in general, slightly lower than that for the radial
oscillations, there is no significant difference between both. The
maximal growth rate for the symmetric case changes weakly with reducing
distance between the loops until the distance becomes comparable to
the radius of the sphere ($z\approx0.5$) (see Figs. \ref{fig: oscZ0}\textit{c}
and \ref{fig: oscZ0}\textit{d} on left). Closer approaching of the
loops causes a significant increase of the growth rate for both axial
and radial oscillations. Again, the smaller the radius of the loops,
the larger the increase of the growth rate. In the antisymmetric case,
the growth rate reaches a maximum as the loops approach to the mid-plane
at a distance comparable to the radius of the sphere ($z\approx1$)
(see Figs. \ref{fig: oscZ0}\textit{c} and \ref{fig: oscZ0}\textit{d}
on right).

\subsection{Spin-up instability}

\begin{figure*}
\centering
\includegraphics[width=\columnwidth]{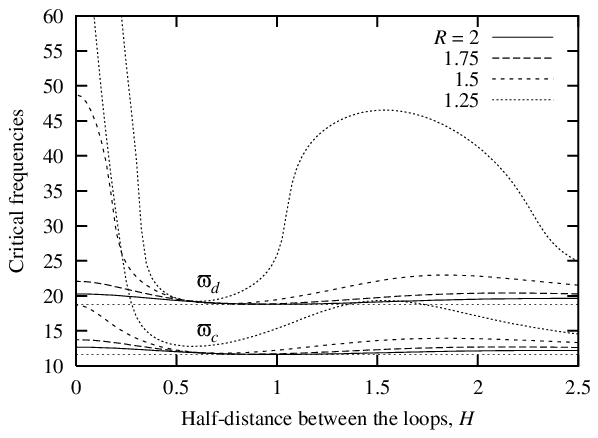}
%{rotcrtZ0s.eps}
{\put(0,0){(a)}}
\includegraphics[width=\columnwidth]{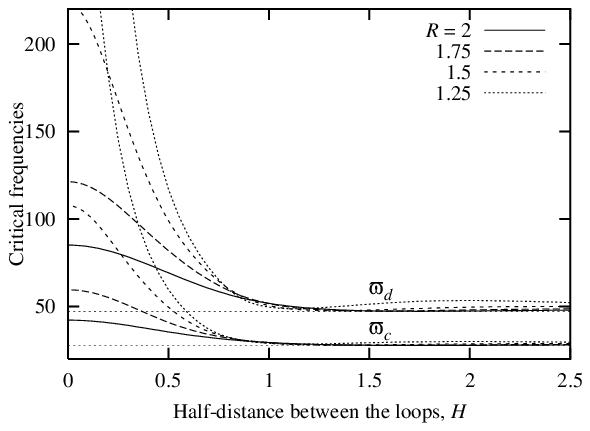}
%{rotcrtZ0a.eps}

\centering
\includegraphics[width=\columnwidth]{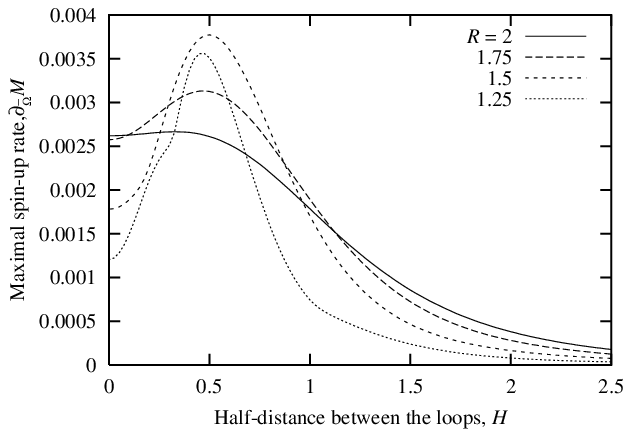}
%{rotmaxZ0s.eps}
{\put(0,0){(b)}}
\includegraphics[width=\columnwidth]{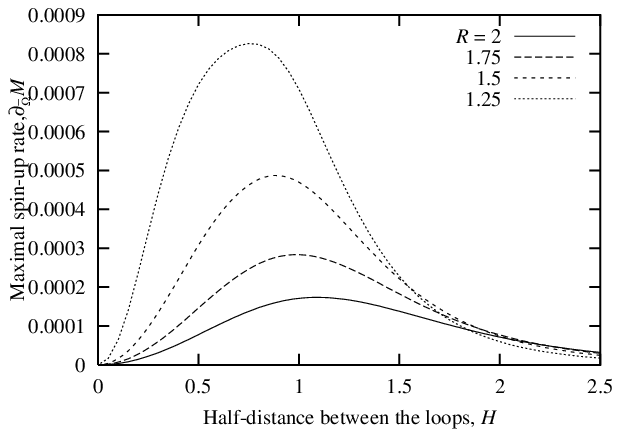}
%{rotmaxZ0a.eps}

\caption{\label{fig: rotZ0} Critical ($\bar{\omega}_{c}$) and most dangerous
($\bar{\omega}_{d}$) dimensionless AC frequencies (\textit{a}) and
maximal growth rates (\textit{b}) for a spin-up instability in symmetric
(on left) and antisymmetric (on right) arrangements of two loops versus
the half-distance of the loops with various radii.}
\end{figure*}

Spin-up instability occurs when the frequency of the alternating magnetic
field exceeds the threshold $\bar{\omega}_{c}$ defined above. The
spin-up rate being proportional to $\partial_{\bar{\Omega}}M$ becomes
positive as the frequency raises over the threshold and attains a
maximum at some higher frequency $\bar{\omega}_{d}$ subsequently
referred to as the most dangerous one. These critical frequencies
together with the corresponding maximal spin-up rates for a sphere
at the middle-point between symmetric and antisymmetric arrangements
of two loops are plotted in Fig. \ref{fig: rotZ0} versus the half-distance
between the loops of various radii. As seen in Fig. \ref{fig: rotZ0}(\textit{a}),
at large enough distances between the loops the threshold and the
most dangerous frequencies tend to $\bar{\omega}_{c}=11.609$ and
$\bar{\omega}_{d}=18.792$ in the symmetric arrangement, and to $\bar{\omega}_{c}=27.682$
and $\bar{\omega}_{d}=47.196$ in the antisymmetric one, which correspond
to the spatially uniform and linear fields, respectively. In the symmetric
case, the critical frequencies have a minimum at the distance between
the loops equal to their radius: $2H=R$ (see Fig. \ref{fig: rotZ0}\textit{a}
on left). Note that this is converse to the critical frequencies for
the oscillatory instability considered previously which have a maximum
at this distance. This minimum is due to the maximal uniformity of
the field in the vicinity of the middle-point achieved at this distance
between the loops. Critical frequencies increase with the non-uniformity
of the field which becomes particularly significant when the loops
are approached closer than the radius of the sphere ($H\approx0.5$).
In this case, the smaller the radius of the loops, the higher the
critical frequencies. The spin-up rate raises as the loops are approached
and attains a maximum at $H\approx0.5$. For the loops with large
enough radius the maximum of the spin-up rate is attained when both
loops merge together forming a single one (see Fig. \ref{fig: rotZ0}\textit{b}
on left). In the antisymmetric case, the critical frequencies begin
to raise significantly when the loops approach to the mid-plane at
a distance of the radius of the sphere: $H\approx1$ (see Fig. \ref{fig: rotZ0}\textit{a}
on right). As seen in Fig. \ref{fig: rotZ0}\textit{b} on right, the
spin-up rate attains a maximum approximately at the same distance.

% Fig. 6

\section{Summary and conclusions}

This work presents an analytic stability analysis of a solid sphere
levitated electromagnetically in an axisymmetric magnetic field induced
by a set of coaxial circular loops. Stability is considered with respect
to both oscillations of small amplitude and arbitrary direction and
rotations perpendicular to the axis of symmetry of the field. Both
oscillations and rotations are found to become growing in time when
the dimensionless frequency of the field exceeds a certain critical
threshold $\bar{\omega}_{c}$ depending on the configuration of the
field for each kind of instability. The growth rates of both instabilities
become positive as the frequency is raised over the corresponding
threshold, attain a maximum at some higher frequency and tend to zero
at high frequencies. The frequency $\bar{\omega}_{d}$, at which the
maximum is attained, is referred to as the most dangerous one for
the corresponding instability. Critical frequencies and the corresponding
maximal growth rates are calculated for arrangements of two loops
of the same radius carrying the same total current which may flow
either in the same or in opposite directions. These arrangements,
corresponding to the usual heating and positioning fields, are referred
to as the symmetric and antisymmetric ones. It is found that critical
frequencies for both oscillatory and rotational instabilities increase
with the nonuniformity of the magnetic field. The lowest dimensionless
critical frequencies are $\bar{\omega}_{c}=11.609$ and $\bar{\omega}_{d}=18.792$
which are the same for both the spin-up instability in a uniform magnetic
field and the oscillatory instability in a spatially linear field.
This coincidence represents a particular case of a more general rule:
the critical frequencies for the spin-up instability in a spherically
harmonic field of degree $l$ coincide with the critical frequencies
for oscillatory instability in a spherically harmonic field of degree
$l+1$. Thus, the critical frequencies $\bar{\omega}_{c}=27.682$
and $\bar{\omega}_{d}=47.196$ for the spin-up instability in a linear
field coincide with the critical frequencies for the oscillatory instability
in a spatially quadratic magnetic field which is the case at the mid-point
of a symmetric arrangement of two loops except the case that the distance
between the loops is equal to the radius of the loops. In this particular
case, the quadratic term vanishes and the instability is dominated
by the fourth-order term which has critical frequencies $\bar{\omega}_{c}=85.252$
and $\bar{\omega}_{d}=158.6$. Thus, the distance $2H=R$ is optimal
in order to avoid the oscillatory instability of the sphere at the
mid-point of the symmetric arrangement. This distance, which ensures
the most uniform magnetic field at the middle-point, yields a minimum
of the critical frequency for the spin-up instability. Note that there
is no similar optimal distance for the oscillatory instability in
antisymmetric arrangement where the critical frequency can be increased
only by the nonuniformity of the magnetic field when the loops are
approached close to the surface of the sphere. Hence, the stabilization
of a sphere by an optimal design of the inductor is rather limited
and active means of stabilization, like an additional steady magnetic
field, may be necessary.

Note that the maximal dimensionless growth rates for both oscillatory
and spin-up instabilities are small, typically $\G\sim10^{-3}$. From
the physical point of view we have introduced the growth rate as a
viscous-type friction coefficient at the velocity. Thus the physical
dimension of $\G$ is $s/m$ and, consequently, its scale is $\tau_{m}/R,$
because we have used the magnetic diffusion time $\tau_{m}$ and the
radius of the sample $R$ as time and length scales, respectively.
In order to obtain the actual friction force we have to take the product
of velocity, friction coefficient and the magnitude of the characteristic
total electromagnetic force $F_{0}$ which usually is comparable to
the gravity of the sample $mg$, where $m$ is the mass of the sample
and $g$ the free fall acceleration. By comparing the negative effective
friction force to the inertia we obtain an estimate of the characteristic
growth time of the instability $\tau_{0}\sim\frac{1}{\G}\frac{R}{g\tau_{m}}$
which for a characteristic size $R\sim10^{-2}~m$ and a conductivity
$\sigma\sim10^{6}~\Omega^{-1}m^{-1}$ of the sample leads to $\tau_{0}\sim10^{4}~s$.
Thus, although the magnetic field has a dynamically destabilizing
effect, the development of the instability is expected to be very
slow and, thus, hardly observable. In conclusion, note that the frequency
of the magnetic field is the only parameter determining the threshold
of dynamic instabilities as long as no external damping, for instance
due to a surrounding gas, is taken into account. In the opposite case,
the threshold of instability would depend not only on the frequency
but also on the amplitude of the current.

In conclusion note that the dynamic instabilities resulting from
the effect of motion of a conducting body in an AC magnetic field
can be interpreted from an alternative physical point of view
which can explain the weakness of this instability and also
suggests other possible instability mechanisms. In the given AC
magnetic field, the electromagnetic force on the spherical body at
rest depends only on its position. If the position changes, it
takes some time for the electromagnetic force to relax to its
time-averaged stationary value for the given position. This delay
is caused by the finite time of magnetic diffusion in the body
which, taking place over the skin depth, is thus comparable to the
AC oscillation period $\tau.$ The electromagnetic force at instant
$t$ on the body in motion may be represented as
$\vec{F}(\vec{r}(t-\tau)).$ Assuming a sufficiently high AC
frequency, we can expand this representation in a power series of
small $\tau$ yielding
$\vec{F}(\vec{r}(t-\tau))\approx\vec{F}(\vec{r}(t)-\tau\vec{v})\approx\vec{F}(\vec{r}(t))-\tau(\vec{v}\cdot\vec{\nabla})\vec{F}.$
Thus, the delay of the magnetic field results in a force component
proportional to the velocity which is obviously analogous to a
viscous friction force. Since for a statically stable position the
effective electromagnetic reaction force
$\vec{(x}\cdot\vec{\nabla})\vec{F}$ is directed against the
displacement $\vec{x}$, the effective friction force
$-\tau(\vec{v}\cdot\vec{\nabla})\vec{F}$ is directed with the
motion. Consequently, the delay of the magnetic field results in
an effective electromagnetic friction force with a negative and,
thus, destabilizing friction coefficient. It is important to note
that the friction coefficient is small because it involves the
delay time which is short and comparable to the AC period when the
relaxation of the magnetic field is solely due to the
electromagnetic diffusion over the skin depth. The same arguments
may be extended to the coupling of the magnetic field and the
temperature distribution in the body caused by its
temperature-dependent electrical conductivity. In this case, the
delay of the magnetic field would be dominated by the thermal
relaxation time which is much longer than the electromagnetic
diffusion time and, thus, expected to cause stronger dynamic
instabilities than the pure electromagnetic mechanism considered
here. In addition, the decrease of the electrical conductivity for
increasing temperature, which is caused by the Joule heating of
the AC magnetic field, and the related reduction of the
non-dimensional frequency $\bar{\omega}$ might be important in
levitation experiments.

\appendices
\section{Approximation of a toroidal inductor of small cross-section
by a circular loop}

Consider a toroidal inductor of arbitrary cross-section carrying
an axisymmetric and purely azimuthal current. The magnetic field
of such an inductor may be represented as a superposition of the
fields of separate circular loops constituting the inductor, and
the corresponding vector potential may be written as the integral
over the inductor cross-section $S$:
$\vec{A}(\vec{r})=\int_{S}\vec{A_{0}}(\vec{r},\vec{r}')ds',$ where
\[
\vec{A_{0}}(\vec{r},\vec{r}')=\frac{1}{4\pi}\int_{0}^{2\pi}
\frac{\vec{j}r'd\phi'}{\left|\vec{r}-\vec{r}'\right|}
\]
is the nondimensionalized vector potential of a single circular
loop defined by the radius vector
$\vec{r}'=r'\vec{e}_{r}'+z'\vec{e}_{z}'$ with $r'$ and $z'$ being
the radius and axial position of the loop in cylindrical
coordinates. Subsequently assume the skin effect to be negligible
as if the inductor would be fed by a direct current driven by the
gradient of an electrostatic potential $\Phi.$ An axisymmetric and
purely azimuthal current distribution with density
$\vec{j}=j\vec{e}_{\phi}$ implies $\Phi$ to depend solely on the
azimuthal angle $\phi$. Consequently,
$\vec{j}=-\vec{\nabla}\Phi=\vec{e}_{\phi}\frac{1}{r}
\frac{\partial\Phi}{\partial\phi},$
where $\frac{\partial\Phi}{\partial\phi}=C$ is a constant that can
be related to the total current in the inductor
$I_{0}=\int_{s}jds$ and its effective radius 
\begin{equation}
\bar{R}=\frac{S}{\int_{S}\frac{ds}{r}}
\label{eq:effr}
\end{equation}
as $C=I_{0}\bar{R}/S.$ Further we assume the inductor cross-section
to be located about some position $\vec{r}_{0}=r_{0}\vec{e}+z_{0}\vec{e}_{z},$ 
which will be specified later, and approximate the field distribution by the
following multipole type expansion:
\setlength{\arraycolsep}{0.0em}
\begin{eqnarray*}
\vec{A}(\vec{r}) &\approx& \frac{I_{0}\bar{R}}{4\pi S}
\int_{0}^{2\pi}\vec{e}_{\phi}'\\
&&{\times}\int_{S}\left[1+(\vec{r}'-\vec{r}_{0})\cdot
\vec{\nabla}_{0}\right]\frac{ds'd\phi'}{\left|\vec{r}-\vec{r}_{0}\right|},
\label{eq:multexp}
\end{eqnarray*}
\setlength{\arraycolsep}{5pt}%
where the operator $\vec{\nabla}_{0}$ acts on $\vec{r}_{0}.$
Now, if we define $\vec{r}_{0}$ as the position of the mass center
of the cross-section $\vec{r}_{0}=\frac{1}{S}\int_{s}\vec{r}ds,$
the second dipole-like term cancels in the expression above which,
thus, reduces to

\begin{equation}
\vec{A}(\vec{r})\approx\frac{\bar{I}r_{0}}{4\pi}\int_{0}^{2\pi}
\frac{\vec{e}_{\phi}'d\phi'}{\left|\vec{r}-\vec{r}_{0}\right|}.
\label{eq:A_approx}
\end{equation}
The last expression defines the vector potential of a circular
loop located at $\vec{r}_{0}$ and carrying the effective current
$\bar{I}=I_{0}\bar{R}/r_{0}.$ For example, in the case of an
inductor represented by a torus with major and minor radii $r_{0}$
and $r_{1},$ the effective radius can easily be found from
(\ref{eq:effr}) as $\bar{R}=\frac{r_{0}}{2}\left(1+\sqrt{1+
\left(\frac{r_{1}}{r_{0}}\right)^{2}}\right)$
while the mass center of the cross-section coincides with its
geometrical center. Note that the solution (\ref{eq:A_approx}) is
accurate up to the quadrapole-like term neglected in
(\ref{eq:multexp}) the magnitude of which relative to the
remaining term may be estimated as $\sim\left|\vec{r}'-
\vec{r}_{0}\right|^{2}/\left|\vec{r}-\vec{r}_{0}\right|^{2}\sim(d/R)^{2},$
where $d$ is the characteristic size of the cross-section and $R$
is the characteristic distance to the cross-section center.
Obviously, the neglected term becomes significant only at
distances comparable to $d.$

\section{Properties of the functions $X_{l}^{m}(\vec{r})$ and
$\bar{X}_{l}^{m}(\vec{r}).$}

Similarly to Ref. \cite{Lohoefer-93} we also use a complex unity
vector $\vec{e}_{\eta}=\frac{1}{\sqrt{2}}\left(\vec{e}_{x}+i\vec{e}_{y}\right),$
where $\vec{e}_{x}$ and $\vec{e}_{y}$ are the corresponding Cartesian
unit vectors, and employ the outer and inner solutions of the scalar
Laplace equation, $X_{l}^{m}(\vec{r})=r^{-l-1}Y_{l,m}(\theta,\phi)$
and $\bar{X}_{l}^{m}(\vec{r})=r^{l}Y_{l,m}(\theta,\phi)$ associated
with the spherical harmonic $Y_{l,m}(\theta,\phi)$ \cite{AbramowitzStegun-65}.
These variables provide a simple algebra for calculation of the gradient
operator $\vec{\nabla}=\vec{e}_{\eta}\frac{\partial~}{\partial\eta^{*}}+
\vec{e}_{\eta}^{*}\frac{\partial~}{\partial\eta}+\vec{e}_{z}
\frac{\partial~}{\partial z}$
used throughout this study: 
\begin{eqnarray*}
\frac{\partial X_{l}^{m}}{\partial\eta} & = & 
\frac{1}{\sqrt{2}}\frac{N_{l+1}^{m-1}}{N_{l}^{m}}X_{l+1}^{m+1},\\
\frac{\partial X_{l}^{m}}{\partial z} & = & 
-\frac{N_{l+1}^{m}}{N_{l}^{m}}X_{l+1}^{m},\\
\frac{\partial\bar{X}_{l}^{m}}{\partial\eta} & = & 
\frac{1}{\sqrt{2}}\frac{2l+1}{2l-1}\frac{N_{l}^{m}}{\bar{N}_{l-1}^{m+1}}
\bar{X}_{l-1}^{m+1},\\
\frac{\partial\bar{X}_{l}^{m}}{\partial z} & = & 
\frac{2l+1}{2l-1}\frac{N_{l}^{m}}{N_{l-1}^{m}}\bar{X}_{l-1}^{m},
\end{eqnarray*}
where the asterisk denotes the complex conjugate and 
$N_{l}^{m}=\sqrt{\frac{(l-m)!(l+m)!}{2l+1}}$.
The corresponding relations for the complex conjugate functions follow
straightforwardly from the above relations and the property of spherical
harmonics: $Y_{l,m}^{*}(\theta,\phi)=(-1)^{m}Y_{l,-m}(\theta,\phi).$
Another relation used for the calculation of the torque in Sec. 3
is: 
\setlength{\arraycolsep}{0.0em}
\begin{eqnarray*}
\vec{r}\times\vec{\nabla}X_{l}^{m} &{}={}& i\left[\frac{\vec{e}_{\eta}^{*}}{\sqrt{2}}
\frac{N_{l}^{m+1}}{N_{l}^{m}}(l-m)X_{l}^{m+1}\right.\\
&&{+}\:\frac{\vec{e}_{\eta}}{\sqrt{2}}\frac{N_{l}^{m-1}}{N_{l}^{m}}(l+m)X_{l}^{m-1}\\
&&\left. {+}\:\vec{e}_{z}mX_{l}^{m}\frac{}{}\right].
\end{eqnarray*}
\setlength{\arraycolsep}{5pt}%
Note that our definition of $X_{l}^{m}$ is slightly different from
that used in \cite{Lohoefer-93} leading to a bit more complicated
algebra but simpler resulting expressions.

\section{Calculation of the coefficients $g_{l}(z).$}

The coefficients $g_{l}(z)$ and $g_{l}'(z)$ in (\ref{eq:F0},\ref{eq:Gamma},
\ref{eq:dM}) can efficiently be calculated using continued fractions \cite{num-recep}.
For this purpose we rewrite $g_{l}(z)=\frac{j_{l+1}(z)}{j_{l-1}(z)}=h_{l}(z)h_{l-1}(z)$
where $h_{l}(z)=\frac{j_{l+1}(z)}{j_{l}(z)}$ and $j_{l}(z)$ is the
spherical Bessel function of index $l.$ Further, applying a recurrence
relation for the spherical Bessel function of index $l$+1 we obtain
the continued fraction: 
\setlength{\arraycolsep}{0.0em}
\begin{eqnarray*}
h_{l}(z) &{}={}& \frac{1}{(2l+3)/z-h_{l+1}(z)}\\
 &{}={}& \frac{1}{(2l+3)/z-}\frac{1}{(2l+5)/z-}\frac{1}{(2l+7)/z-}...
\end{eqnarray*}
\setlength{\arraycolsep}{5pt}
allowing us to calculate $h_{l}(z)$ provided that $h_{l+1}(z)$
is known. But asymptotic properties of Bessel functions suggest that
for large index $l$ $h_{l}(z)\sim\frac{z}{2l}$. Thus for any $z$
we can choose sufficiently large $l'$ and then truncate the fraction
by approximating $h_{l'}(z)$ by the previous expression that allows
us to calculate back the necessary $h_{l}(z).$ Note that the forward
recurrence for $h_{l}(z)$ is not practically applicable because it
is numerically unstable similarly to its counterpart for Bessel functions.
The other necessary quantity $g_{l}'(z)=\frac{dg_{l}}{dz},$ contained
in (\ref{eq:Gamma},\ref{eq:dM}), can be calculated in a similar
way by expressing it as $g_{l}'(z)=(2l+1)(h_{l-1}^{2}(z)-g_{l}(z))/z$.
For $\bar{\omega}\gg1$ we have $g_{l}(\sqrt{\bar{\omega}/i})
\sim-1+\frac{2l+1}{\sqrt{i\bar{\omega}}}$ which follows from the 
corresponding asymptotics of the Bessel functions \cite{AbramowitzStegun-65}.

\end{document}